\documentclass[12pt]{article}
\pdfoutput=1

\usepackage[usenames,dvipsnames]{xcolor}
\usepackage{tikz}
\usetikzlibrary{matrix,arrows,positioning,shapes,snakes,fadings,decorations.pathmorphing,
decorations.pathreplacing,automata,shadows,decorations.markings,patterns,decorations.text}
\usepackage{jheppub}
\usepackage{amsmath,amssymb,euscript,array,mathrsfs,appendix,ctable,mathabx}
\usepackage{mathtools,float}
\usepackage{arydshln}
\usepackage{todonotes}
\usepackage{graphicx}
\usepackage{wrapfig}
\usepackage{enumitem}
\usepackage{multirow}
\usepackage{bm,upgreek}
\usepackage{stmaryrd}
\usepackage{verbatim}

\usepackage{caption}
\usepackage{subcaption}

\usepackage{amsthm}

\makeatletter
\newsavebox{\@brx}
\newcommand{\llangle}[1][]{\savebox{\@brx}{\(\m@th{#1\langle}\)}%
  \mathopen{\copy\@brx\kern-0.6\wd\@brx\usebox{\@brx}}}
\newcommand{\rrangle}[1][]{\savebox{\@brx}{\(\m@th{#1\rangle}\)}%
  \mathclose{\copy\@brx\kern-0.6\wd\@brx\usebox{\@brx}}}
\makeatother

\newcommand*\bigcircled[1]{{\tikz[baseline=(char.base)]{%
            \node[shape=circle, fill=black!20,draw,inner sep=1pt] (char) {#1};}}}

\newcommand \arXiv [1]{\href{http://arxiv.org/abs/#1}{\tt arXiv:#1}} 
 \newcommand{\bea}{\begin{eqnarray}}
\newcommand{\eea}{\end{eqnarray}}
\newcommand{\be}{\begin{equation}}
\newcommand{\ee}{\end{equation}}

\usepackage{titlesec}
\titleformat{\subsection}[display]{\it}{}{0.1cm}{\vspace{-1.5cm}\begin{center}\thesubsection\hspace{0.2cm}}[\end{center}\vspace{-0.5cm}]

\newcommand\MAT[1]{\begin{pmatrix} #1\end{pmatrix}}

\newcommand{\EQ}[1]{\begin{equation}\begin{split} #1
\end{split}\end{equation}}

\title{Replica R\'enyi Wormholes and Generalised Modular Entropy in JT Gravity}
\author{Timothy J. Hollowood, S. Prem Kumar and Luke C. Piper}
\affiliation{Department of Physics, Swansea University, Swansea, SA2 8PP, U.K.}
\emailAdd{t.hollowood@swansea.ac.uk, s.p.kumar@swansea.ac.uk, l.c.piper.2117970@swansea.ac.uk}
\abstract{
We consider the problem of computing semi-classical R\'enyi entropies of  CFT on  AdS$_2$ backgrounds in JT gravity with nongravitating baths, for general replica number $n$. Away from the $n\to 1$ limit,  the backreaction of the CFT twist fields on the geometry is nontrivial. For one twist field insertion and general $n$, we show that the quantum extremal surface (QES) condition involves extremisation of the {\em generalised  modular entropy}, consistent with Dong's generalisation of the Ryu-Takayanagi formula for general $n$. For multiple QES we describe  replica wormhole geometries using the theory of Fuchsian uniformisation, explicitly working out the analytically tractable case of the  $n=2$ double trumpet wormhole geometry. We determine the off-shell dependence of the gravitational action on the QES locations and boundary map. In a factorisation limit, corresponding to late times, we are able to relate this action functional to area terms given by the value of the JT dilaton at the (off-shell) QES locations, with computable corrections. Applied to the two-sided eternal black hole, we find the $n$-dependent Page times for R\'enyi enropies in the high temperature limit.
}

\notoc
\begin{document}

\maketitle

\newpage 

%\tableofcontents 

\section{Introduction}

The celebrated island formula \cite{Almheiri:2019hni} provides a semiclassical window into the resolution of Hawking's long-standing information puzzle for black holes \cite{Hawking:1976ra}.  Lying at the heart of this mechanism  and the associated generalised entropy formula \cite{Almheiri:2019psf, Penington:2019npb, Engelhardt:2014gca, Faulkner:2013ana, Hubeny:2007xt}, is the replica trick for dynamical gravity. The gravitational replica trick necessitates the inclusion of replica wormhole saddle points  in the  sum over geometries implemented by the gravitational path integral \cite{Almheiri:2019qdq, Penington:2019kki}.  This yields the island formula\footnote{The island formula and its implications have been explored in a variety of different scenarios and involving evaporating black holes } for the {\it generalised} von Neumann entropy of Hawking radiation, a fine-grained measure\footnote{The notion of a fine-grained entropy here is a relative one, as emphasised in \cite{Ghosh:2021axl, Krishnan:2020oun}.} of entanglement which teases out subtle quantum correlations between early and late radiation, and reproduces the Page curve \cite{Page:1993wv}.

It is worth stressing that the generalised entropy formula 
%for Hawking radiation 
%$R$,
%\begin{equation}
%S_{\rm gen}(R) = {\rm min}\,{\rm ext}_{\{I\}}\left[\frac{{\rm Area}(\partial I)}{4 G_N}+ S_{\rm QFT}(I\cup R)\right]\,,
%\end{equation}
is an off-shell expression for the fine-grained entropy of Hawking radiation $R$. Obtained in  the formal limit when the number of replicas $n\to 1$,  it consists   of an area term evaluated at a Quantum Extremal Surface (QES) demarcating the boundary of a putative island region $I$, together with a QFT contribution to entanglement entropy of $I\cup R$. The implication is that the island is in the Hilbert space of the radiation,  or more precisely, $I$ lies  in the entanglement wedge of the radiation $R$.  

Our motivation is to understand  if a similar generalised formula exists for the R\'enyi entropies $\{S_n\} $ \cite{renyi1}, for general replica number $n$, not reliant on the crutch of the $n\to 1$ limit. R\'enyi entropies are measurable \cite{Islam:2015mom},  encoding the quantum state and entanglement spectrum of the subsystem.  Our main observation is that there {\em is} a generalised R\'enyi entropy formula within the Euclidean framework which, in late time or factorization limits, has the same structure as the generalised entanglement entropy, and where the area term involves not the R\'enyi entropy, but the closely related `modular entropy' discovered by Dong in the setting of holography \cite{Dong:2016fnf}.

The semiclassical derivation of the island formula is most explicit  within the framework of Jackiw-Teitelboim (JT) gravity \cite{Jackiw:1984je, Teitelboim:1983ux} which naturally arises as an effective 1+1 dimensional description of gravity, following from reduction to $s$-waves around a higher dimensional near extremal black hole.  JT gravity has proved to be a very useful model to explore multiple aspects of  information  evolution and retrieval from evaporating black holes
\cite{Almheiri:2019yqk,Almheiri:2019qdq,Penington:2019kki,Goto:2020wnk}. 

In particular, the saddles that contribute to computation of the matter R\'enyi entropies, the so-called replica wormholes, have been constructed in the pioneering works \cite{Penington:2019npb,Almheiri:2019psf,Almheiri:2019qdq,Penington:2019kki}, at least in certain regimes and this allows one to confirm that the entropy of the radiation follows the Page curve. Specifically, \cite{Almheiri:2019qdq} considers a setup originally introduced in \cite{Almheiri:2019yqk}, involving the eternal AdS$_2$ black hole in JT gravity coupled to a CFT playing the r\^ole of a radiation bath.  Most attention has focussed around the computation of  von Neumann entropies for radiation subsystems in this framework. This involves taking a limit $n\to1$ of the replica wormholes, which makes the problem much simpler because  the gravitational back-reaction of the CFT twist operators needed to define the R\'enyi entropy vanishes in this limit. What remains is a variational problem for the positions of the twist operators, leading to the so-called Quantum Extremal Surface (QES) formula. Remarkably this yields an explicit  expression for the microscopic, or fine-grained entropy of the radiation in a calculation that is purely semi-classical. 

In this work, we set out to address the question whether one can write down a variational problem for  R\'enyi entropies  that generalizes the well-known QES formula for von Neumann entropy. To this end, we consider replica wormholes within the above setup when the limit $n\to1$ is not taken and the back-reaction of  twist operators must be taken into account when constructing replica wormholes. Previous work in this direction for JT gravity appears in \cite{Colin-Ellerin:2021jev}. 

The second motivation for this work arose from a potential puzzle. The equations of motion for JT gravity (in particular, the Einstein equations) coupled to a CFT radiation bath, show how the dilaton field $\phi$  in JT gravity, which plays the r\^ole of the area in this two-dimensional effective theory of gravity, gets sourced by the CFT stress tensor. For any number of replicas $n$, this equation, along with CFT Ward identities leads to a set of conditions that determine the positions of the twist fields $w=a_j$, i.e.~the QES, 
\EQ{
\frac{\partial_w\phi(a_j,\bar a_j)}{4G_N}+\partial_{a_j} S^\text{CFT}_n=0\ ,
\label{wii}
}
where $S_n^\text{CFT}$ is the R\'enyi entropy of the matter CFT. This equation was inferred in  \cite{Almheiri:2019qdq} and then shown to arise from a simple argument involving the dilaton equation of motion and the CFT Ward identities in \cite{Goto:2020wnk}. 

Whilst tempting to argue that  \eqref{wii} is already the general $n$ QES formula that we are seeking, closer inspection reveals two important distinctions:
\begin{enumerate}
\item{Since the dilaton plays the r\^ole of the area, in \eqref{wii} it is expected to be  the JT gravity version of the Ryu-Takayanagi (RT) formula \cite{Ryu:2006bv} in holography.\footnote{This is exactly what one expects when one views the JT gravity black hole as describing the s-wave sector of a near-extremal charged black hole in $3+1$ dimensions.} There is already a generalization of the RT formula to R\'enyi entropy discovered by Dong \cite{Dong:2016fnf} (see also \cite{Dong:2017xht,Dong:2018seb}). In particular, Dong argued that the RT term would naturally be associated not to the R\'enyi entropy but to a close refinement  called the `modular entropy'. The modular entropy $\widetilde S_n$ is the natural thermodynamical definition of the entropy when the replica number $n$ is identified with the inverse temperature  and the free energy $F_n= -\frac1n\log Z_n^{\rm CFT}$ where $Z_n^{\rm CFT}$ is the replica partition function,
\be
\widetilde S_n= -\partial_{\frac1n}F_n\,=\,n^2\partial_n\Big(\frac{n-1}nS_n^{\rm CFT}\Big)\,.
\ee 
For a single interval, wherein CFT R\'enyi entropies are universal, the modular entropy has an $n$ dependence distinct from the R\'enyi entropy, although they are then closely related:
\be
\widetilde S_n^\text{CFT}=\frac2{n+1}S_n^\text{CFT}=\frac1n {S_{n=1}^\text{CFT}}\,.
%\qquad S_n^{\rm CFT} =\frac{n+1}{2n}\,{S^\text{CFT}}
\ee
}
\item{ The well known QES formula for entanglement entropy results from extremising the generalised entropy which includes the variation of the area term (the dilaton) with respect to the off-shell positions of punctures $\sim \partial_{a_j}\phi(a_j)$, whereas \eqref{wii} involves the derivative of the area term evaluated at the punctures $\sim\partial_w\phi\big|_{w=a_j}$. }
\end{enumerate}
%  If one takes this view, then  defined by in terms of the R\'enyi entropy via
%\EQ{
%\widetilde S_n=n^2\partial_n\Big(\frac{n-1}nS_n\Big)\ .
%}
The two observations above put together suggest that the QES equation for $n>1$ should be of the form
\EQ{
\frac{\partial_{a_j}\phi(a_j,\bar a_j)}{4G_N}+\partial_{a_j}\widetilde S^\text{CFT}_n=0\ ,
\label{wii2}
}
extremising a generalised modular entropy, $\widetilde S^{\rm gen}_n$. 
The puzzle is that, taken at face value the form of \eqref{wii2} appears to have the wrong $n$ dependence to match the Ward identity equation \eqref{wii}.  For multiple intervals the modular and R\'enyi entropies differ nontrivially and the puzzle is more acute. The second item above provides the  potential loophole to make the mystery evaporate.
We will argue that whilst equation \eqref{wii} is certainly correct, \eqref{wii2} is also true. When there are multiple QES, this will be true in a certain factorization limit.
%Even the single interval results for  in a CFT on the plane for which the resuare universal, the modular and R\'enyi and von Neumann entropies are all simply related:
%\EQ{
%\widetilde S_n^\text{CFT}=\frac2{n+1}\,S_n^\text{CFT},\qquad S_n^{\rm CFT} =\frac{n+1}{2n}\,{S^\text{CFT}}\ .
%}
%The puzzle is that the form of \eqref{wii2} seems to have the wrong $n$ dependence to match the Ward identity equation \eqref{wii}. 
%If one looks carefully at the two equations \eqref{wii} and \eqref{wii2} the mystery evaporates and a loop hole emerges. The Ward identity equation involves the $w$ derivative of the dilaton evaluated at the QES whilst the RT inspired equation \eqref{wii2} involves the derivative with respect to the position of the QES of the dilaton evaluated at the QES. 
For $n>1$ there is non-trivial gravitational backreaction from the twist fields on the geometry and in JT gravity this means that the dilaton evaluated at the QES will have both explicit {\it and\/} implicit dependence on the QES.

We will employ uniformisation techniques to compute the replica partition function of the gravity plus CFT system as a function of the off-shell QES locations  and the boundary map which glues the CFT radiation baths to the gravitating  AdS$_2$  region. For a single QES in the eternal black hole setup, we can explicitly demonstrate the validity of \eqref{wii2} for any $n$. Computing the on-shell value of the generalised modular entropy for generic  $n>1$ is technically challenging, as it requires knowing the boundary map precisely.

For multiple QES, we show that there is a generalized R\'enyi entropy which has the approximate form 
\EQ{
S_n(a_j,\bar a_j)=NS_0+\frac{n+1}{8G_N}\sum_j\Big(\phi_j(a_j,\bar a_j)+\Phi_n\Big)+S_n^\text{CFT}(a_j,\bar a_j)\ .
}
The piece $\Phi_n$ can be interpreted as the effect of back-reaction on a particular QES from all the other QES. This term comes with a factor of $n-1$ and so does not contribute when taking the $n\to1$ limit. Fortunately, it is suppressed in the factorization limit that is relevant for understanding the transition from the saddle without an island (the Hawking saddle) and the saddle with an island (replica wormhole) that happens at late times, precisely when the replica wormhole saddle can dominate. With this term neglected and with knowledge that $S_n^\text{CFT}=\frac{n+1}2\widetilde S_n^\text{CFT}$, we see that the QES equation that follows from extremizing the above is precisely of the form \eqref{wii2}. In addition, since the $n$ dependence of the dilaton and the modular entropy match, i.e.~go like $1/n$, means that the position of the QES is independent of $n$ in the high-temperature, late-time, regime. This simple picture is valid for the two-sided external black hole which is in thermal equilibrium with the radiation baths. It is not so simple for the case of an evaporating black hole which will be considered in a separate work. The difficulty is that the equation of motion for the boundary map is too complicated to solve in general (along with the associated conformal welding problem \cite{Almheiri:2019qdq,Goto:2020wnk}) away from the $n\to1$ limit.  For the set up involving the eternal black hole, there is a simplification to be exploited  in the high temperature regime  wherein one can use perturbation theory to argue that the boundary map is only a small perturbation of the trivial map \cite{Almheiri:2019qdq}. 

The paper is organized as follows. In section \ref{s1.5}, we review some relevant aspects of JT gravity and replica wormholes. In section \ref{s2}, we consider the case with one QES which is non-trivial but also  tractable because we do not have to consider the effects of back-reaction of one QES on another. We apply the result to 
analyse the eternal black hole scenario in the high temperature limit. Section \ref{s3} reviews aspects of the theory of uniformisation relevant for the construction  of the replica wormholes, which we make very explicit for the case of $n=2$ and with two QES. We also exploit the well-known relation of the uniformisation problem to Liouville theory. In section \ref{s4}, we turn to an analysis of the dilaton in the replica wormhole geometry using the Green's function method generalized from the $n=1$ case. This gives us enough information to compute the dominant terms in the gravitational action in the factorisation limit, and apply the results to compute the Page times for R\'enyi entropies for the two sided black hole. In section \ref{s5}, we bring together our results to draw some conclusions and outline immediate future directions.

\section{JT gravity and replica wormholes}\label{s1.5}

The basic setup used to study black hole evaporation in JT gravity involves coupling the gravitational system to an external non-gravitating Minkowski bath. In the semiclassical picture, massless degrees of freedom or CFT modes  propagate on the gravitational background and the Minkowski bath across a transparent bath-gravity interface.\footnote{One can, in principle, turn on non-vanishing reflection coefficients at the interface to capture the effects of black hole greybody factors \cite{Kruthoff:2021vgv}.} The backreaction of the CFT on gravity is treated semiclassically, so the central charge $c$ is assumed to be large.

Although black hole evaporation and associated entanglement evolution is intrinsically a Lorentzian phenomenon, the Euclidean continuation of the setup described above is relevant and particularly useful when considering entanglement dynamics of the two-sided black hole in the thermofield double (TFD) state. Whilst the system is in {\em local} thermodynamic equilibrium, we expect the time evolution of quantum correlations between  radiation modes across the two copies to be captured by the Euclidean setting accompanied by appropriate analytic continuation .

So we focus attention on the Euclidean section of the AdS$_2$ black hole (the disc). The equation of motion for the dilaton in JT gravity constrains the Ricci scalar $R=-2$, and therefore fixes the metric to be locally AdS$_2$.
The Euclidean CFT lives on the gravitating disc\footnote{Since the metric is fixed to be locally AdS$_2$ the dynamical degrees of freedom are the dilaton and the boundary curve of the disc.} glued on to the flat non-gravitating plane. We work initially in Euclidean signature with complex conjugate coordinates $(w,\bar w)$ for the locally AdS$_2$ gravitating region and  $(y,\bar y)$ for the flat bath: 
\bea
&&ds^2=e^{2\rho(w,\bar w)}\,|dw|^2\,,\qquad |w|\leq 1\,,\\\nonumber\\
&&ds^2_{\rm bath} =  |dy|^2\,,\qquad {\rm Re}(y) \geq 0\,,
\eea
with $y=\sigma + i\tau$, and $\rho(w,\bar w)$ the conformal factor to be determined by the equations of motion. The analytic continuation to Lorentzian signature via the Wick rotation $\tau\to it$ involves $w\to -w^-$ and $\bar w\to w^+$ and similarly $y\to -y^-$ and $\bar y\to y^+$ where $y^\pm=t\pm\sigma$ are coordinates in the Minkowski baths. The gluing map between the boundaries of these two regions plays an important dynamical role below.

\subsection{Replica geometry}
\begin{figure}[ht]
\begin{center}
\begin{tikzpicture} [scale=0.7,every node/.style={scale=0.8},decoration={markings,mark=at position 0.5 with {\arrow{>}}}]]
\node at (-2,9.5) {$a_2$};
\node at (-5.5,9.5) {$1/\bar a_2$};
\node at (2,9.5) {$a_1$};
\node at (5.5,9.5) {$1/\bar a_1$};
\node at (1,7.9) {\bigcircled{3}};
\node at (1,4.9) {\bigcircled{2}};
\node at (1,1.9) {\bigcircled{1}};
\draw[very thick,fill=black!30,opacity=0.2] (0,2) ellipse (4cm and 1cm);
\draw[very thick,fill=black!30,opacity=0.2] (0,5) ellipse (4cm and 1cm);
\draw[very thick,fill=black!30,opacity=0.2] (0,8) ellipse (4cm and 1cm);
\draw[thick,dotted,fill=blue!20,opacity=0.2] (-2,8) -- (-5.5,8) -- (-5.5,2) -- (-2,2) --cycle;
\draw[thick,dotted,fill=blue!20,opacity=0.2] (2,8) -- (5.5,8) -- (5.5,2) -- (2,2) --cycle;
\draw[decorate,very thick,decoration={snake,segment length=2mm,amplitude=0.4mm},blue] (-2,7.93) -- (-5.5,7.93);
\draw[decorate,very thick,decoration={snake,segment length=2mm,amplitude=0.4mm},green] (-2,8.07) -- (-5.5,8.07);
\draw[decorate,very thick,decoration={snake,segment length=2mm,amplitude=0.4mm},green] (-2,4.93) -- (-5.5,4.93);
\draw[decorate,very thick,decoration={snake,segment length=2mm,amplitude=0.4mm},red] (-2,5.07) -- (-5.5,5.07);
\draw[decorate,very thick,decoration={snake,segment length=2mm,amplitude=0.4mm},red] (-2,1.93) -- (-5.5,1.93);
\draw[decorate,very thick,decoration={snake,segment length=2mm,amplitude=0.4mm},blue] (-2,2.07) -- (-5.5,2.07);
\draw[decorate,very thick,decoration={snake,segment length=2mm,amplitude=0.4mm}] (5.5,8) -- (2,8);
\draw[decorate,very thick,decoration={snake,segment length=2mm,amplitude=0.4mm}] (5.5,5) -- (2,5);
\draw[decorate,very thick,decoration={snake,segment length=2mm,amplitude=0.4mm}] (5.5,2) -- (2,2);
\filldraw[red] (2,8) circle (0.1cm);
\filldraw[red] (2,5) circle (0.1cm);
\filldraw[red] (2,2) circle (0.1cm);
\filldraw[black] (-2,8) circle (0.1cm);
\filldraw[black] (-2,5) circle (0.1cm);
\filldraw[black] (-2,2) circle (0.1cm);
\filldraw[black] (-5.5,8) circle (0.1cm);
\filldraw[black] (-5.5,5) circle (0.1cm);
\filldraw[black] (-5.5,2) circle (0.1cm);
\filldraw[black] (5.5,8) circle (0.1cm);
\filldraw[black] (5.5,5) circle (0.1cm);
\filldraw[black] (5.5,2) circle (0.1cm);
\draw[thick,blue] (0,2.7) .. controls (-1,2.7) and (-3,2.5) .. (-3,2.1);
\draw[thick,blue] (3,8) .. controls (3,7.1) and (-3,7.1) .. (-3,8);
\draw[thick,blue,<-] (0.2,2.7) .. controls (1,2.7) and (3,2.5) .. (3,2);
\end{tikzpicture}
\caption{\footnotesize The shaded part is the replica wormhole $\widetilde{\cal M}_n$ with $N=2$ and $n=3$ so a triple cover of the base the twice punctured disc $|w|\leq1$. The sheets are joined by branch cuts between the QES as shown involving image branch points outside the disc. The cuts are colour-coded so that one can identify their image in the uniformized picture in figure \ref{fig2}.}
\label{fig1} 
\end{center}
\end{figure}
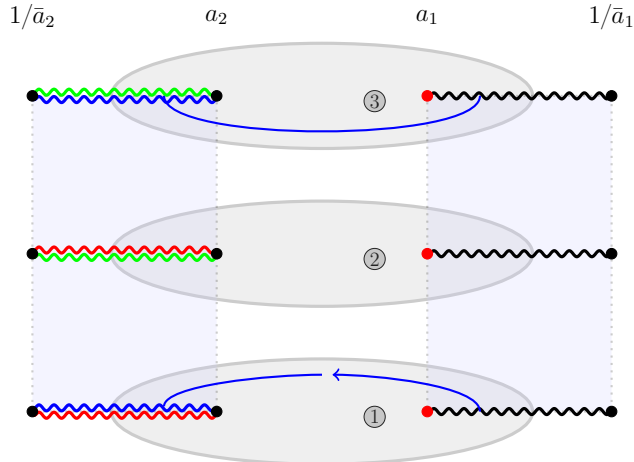

When computing a CFT R\'enyi entropy one can either proceed via the replica method sewing together $n$ copies of the spacetime along a set of branch cuts \cite{Calabrese:2004eu}, or by computing a correlator of twist and anti-twist operators in the un-replicated geometry at the positions of the branch points. In the gravitating AdS$_2$ region in JT gravity we can also consider the problem either in the replicated geometry $\widetilde{\cal M}_n$ or in the `base' ${\cal M}_n$ with the twist operators. The replicated geometry  $\widetilde{\cal M}_n$ can be thought of as a Riemann surface that is an $n$-fold cover of the base with respect to the assumed ${\mathbb Z}_n$ replica symmetry, ${\cal M}_n=\widetilde{\cal M}_n/{\mathbb Z}_n$.  The base is the unit disc $|w|\leq1$ with $N$ punctures corresponding to branch points $\{a_j\}_{j=1,2\ldots N}$ that link the $n$ sheets of the cover. These branch points will become the QES after solving a variational problem. The branch cuts originating from each of the QESs end at an image point outside the disc, in the bath.The resulting Riemann surface has genus $g=(n-1)(N-1)$. The boundary $\partial\widetilde{\cal M}_n$ is a single $S^1$, for $N$ odd, and $(S^1)^n$, for $N$ even. An example, with $n=3$ and $N=2$ QES is shown in figure \ref{fig1}.

The metric in the covering space is  the AdS$_2$ metric written in standard form
\EQ{
ds^2=\frac{4|dW|^2}{(1-|W|^2)^2}\ .
\label{zaz}
}
The uniformisation map $\pi:\ \widetilde{\cal M}_n\to{\cal M}_n$ maps the replica geometry to the base ${\cal M}_n$, the unit disc $|w|\leq1$. The multi-valued inverse map $W(w)$ determines the metric on the base:
\EQ{
ds^2=e^{2\rho}\,|dw|_n^2\ ,\qquad e^{2\rho}=\frac{4 |\partial_wW(w)|^2}{(1-|W(w)|^2)^2}\ ,
\label{pmm}
}
which is single-valued on the base. Near an insertion point of a twist  field at $w=a_j$, the inverse map 
\be
W(w) \sim (w-a_j)^{\pm 1/n}+\ldots,
\ee
 determines the branch point singularity of the conformal factor $e^{2\rho}$, the $+$ and $-$ signs in the exponent associated to twist and anti-twist fields respectively. Taking the viewpoint of the base ${\cal M}_n$, the twist operators make punctures in the geometry with conical  deficits with angles $2\pi(1-\frac1n)$. These deficits (for both twist and anti-twist fields) are manifested in the equation for the conformal factor of the metric $ds^2=e^{2\rho}\,|dw|_n^2$ as source terms  \cite{Almheiri:2019qdq},
\EQ{
-4\partial_w\partial_{\bar w}\rho+e^{2\rho}=2\pi\Big(1-\frac1n\Big)\sum_j\delta^{(2)}(w-a_j)\ .
\label{ldr}
}
In the absence of source terms, the equation simply sets the Ricci scalar $R=-2$, so the metric is clearly locally AdS$_2$. The $|dw|_n^2$ part  is the metric on the punctured disc with angular deficits of $2\pi(n-1)/n$ at each puncture so that the metric on the cover \eqref{zaz} is smooth. The conformal factor approaches that of AdS near the boundary so that
\EQ{
e^{-2\rho}=(1-|w|^2)^2+{\mathscr O}((1-|w|^2)^4)\ .
\label{dew}
}
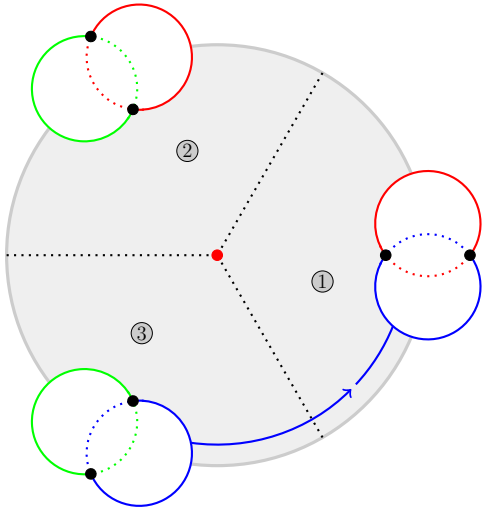
\begin{figure}[ht]
\begin{center}
\begin{tikzpicture} [scale=0.7,every node/.style={scale=0.8},decoration={markings,mark=at position 0.5 with {\arrow{>}}}]]
\draw[very thick,fill=black!30,opacity=0.2] (0,0) circle (4cm);
\draw[white,fill=white] (4,0.6) circle (1cm);
\draw[white,fill=white] (4,-0.6) circle (1cm);
\draw[thick,red] ([shift=(-32:1cm)]4,0.6) arc (-32:212:1cm);
\draw[thick,red,dotted] ([shift=(-32:1cm)]4,0.6) arc (-32:-158:1cm);
\draw[thick,blue] ([shift=(32:1cm)]4,-0.6) arc (32:-212:1cm);
\draw[thick,blue,dotted] ([shift=(32:1cm)]4,-0.6) arc (32:158:1cm);
\filldraw[black] (3.2,0) circle (0.1cm);
\filldraw[black] (4.8,0) circle (0.1cm);
\draw[dotted,thick] (-4,0) -- (0,0);
\node at (2,-0.5) {\bigcircled{1}};
\begin{scope}[rotate=120]
\draw[white,fill=white] (4,0.6) circle (1cm);
\draw[white,fill=white] (4,-0.6) circle (1cm);
\draw[thick,green] ([shift=(-32:1cm)]4,0.6) arc (-32:212:1cm);
\draw[thick,green,dotted] ([shift=(-32:1cm)]4,0.6) arc (-32:-158:1cm);
\draw[thick,red] ([shift=(32:1cm)]4,-0.6) arc (32:-212:1cm);
\draw[thick,red,dotted] ([shift=(32:1cm)]4,-0.6) arc (32:158:1cm);
\filldraw[black] (3.2,0) circle (0.1cm);
\filldraw[black] (4.8,0) circle (0.1cm);
\draw[dotted,thick] (-4,0) -- (0,0);
\node at (2,-0.5) {\bigcircled{2}};
\end{scope}
\begin{scope}[rotate=-120]
\draw[white,fill=white] (4,0.6) circle (1cm);
\draw[white,fill=white] (4,-0.6) circle (1cm);
\draw[thick,blue] ([shift=(-32:1cm)]4,0.6) arc (-32:212:1cm);
\draw[thick,blue,dotted] ([shift=(-32:1cm)]4,0.6) arc (-32:-158:1cm);
\draw[thick,green] ([shift=(32:1cm)]4,-0.6) arc (32:-212:1cm);
\draw[thick,green,dotted] ([shift=(32:1cm)]4,-0.6) arc (32:158:1cm);%
\filldraw[black] (3.2,0) circle (0.1cm);
\filldraw[black] (4.8,0) circle (0.1cm);
\draw[dotted,thick] (-4,0) -- (0,0);
\node at (2,-0.5) {\bigcircled{3}};
\end{scope}%
\filldraw[red] (0,0) circle (0.1cm);
%
%\draw[postaction={decorate},*-*] (2.9,1.9) to[out=-60,in=90] (3.4,0.2);
%\draw[postaction={decorate},*-*] (2.8,1.3) to[out=130,in=20] (-2,2.5);
%\draw[postaction={decorate},*-*] (1,0) arc[start angle=0,end angle=120,radius=1cm];
%
%\node at (0.3,0.4) {$M_1$};
%\node at (3.9,1.6) {$M_2$};
%\node at (0,3.3) {$M_2M_1$};
%
%
\draw[->,thick,blue] ([shift=(-98:3.6cm)]0,0) arc (-98:-45:3.6cm);
\draw[thick,blue] ([shift=(-43:3.6cm)]0,0) arc (-43:-22:3.6cm);
\end{tikzpicture}
\caption{\footnotesize The uniformisation view of the replica wormhole in figure \ref{fig1}. The fundamental domain ${\cal D}$ is the shaded region, i.e.~the disc with the circles cut out. The coloured arcs of the circles are then identified and these correspond to how the cuts are joined in figure \ref{fig1} and gives rise to the illustration of the replica wormhole in figure \ref{fig3}.}
\label{fig2} 
\end{center}
\end{figure}
 The replicated geometry $\widetilde{\cal M}_n$ is most conveniently described by using the theory of uniformisation, in this context Fuchsian uniformisation \cite{Hulik:2016ifr,Hadasz:2005gk,Hadasz:2006rb}. Essentially $\widetilde{\cal M}_n$ is covered by a complex coordinate $W$ defined on a domain ${\cal D}$ that is a subset of the unit disc $|W|\leq1$ formed by cutting out a series of circles that straddle the boundary $|W|=1$. Details will be given later on how the circles are defined but an illustration for the $n=3$ and $N=2$ example is shown in figure \ref{fig2}. The coloured arcs on the circles are identified so that the closed blue contour in figure \ref{fig1} whose image is shown in figure \ref{fig2} is actually closed modulo the non-trivial identifications in the covering space. Joining up the contours gives a picture of the replica wormhole as a smooth Riemann surface as shown in figure \ref{fig3}.

\begin{figure}[ht]
\begin{center}
\begin{tikzpicture} [scale=0.8] 
\draw[very thick,black!30] (0,3) ellipse (1.5cm and 0.5cm);
\begin{scope}[rotate=120]
\draw[very thick,black!30] (0,3) ellipse (1.5cm and 0.5cm);
\end{scope}
\begin{scope}[rotate=-120]
\draw[very thick,black!30] (0,3) ellipse (1.5cm and 0.5cm);
\end{scope}
\draw[very thick,black!30] (-1.5,3) to[out=-90,in=40] (-3.3,-0.198);
\draw[very thick,black!30,rotate=120] (-1.5,3) to[out=-90,in=40] (-3.3,-0.2);
\draw[very thick,black!30,rotate=-120] (-1.5,3) to[out=-90,in=40] (-3.3,-0.2);
\draw[very thick,blue] (0.2,-0.2) to[out=-30,in=155] (3,-1.8);
\draw[very thick,red] (0.2,-0.2) to[out=100,in=-90] (0,2.5);
\draw[very thick,green] (0.2,-0.2) to[out=-150,in=20] (-2.1,-1.3);
\draw[thick,dotted] (-0.5,0.5) to[out=-30,in=155] (2.2,-1.1);
\draw[thick,dotted] (-0.5,0.5) to[out=90,in=-90] (-0.5,3.5);
\draw[thick,dotted] (-0.5,0.5) to[out=-150,in=25] (-3.2,-1.4);
\filldraw[red] (-0.5,0.5) circle (0.1cm);
\filldraw[black] (0.2,-0.2) circle (0.1cm);
\end{tikzpicture}
\caption{\footnotesize The replica wormhole of figures \ref{fig1} and \ref{fig2} pictured as a smooth Riemann surface showing the two branch points, the QES, and the images of the coloured cuts.}
\label{fig3} 
\end{center}
\end{figure}
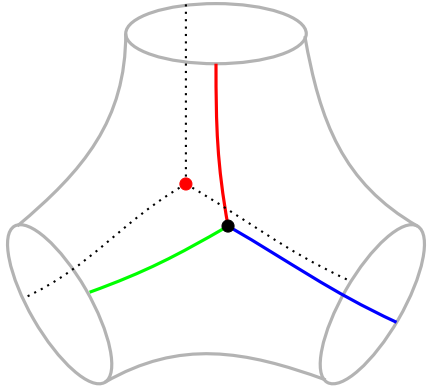

\subsection{The dilaton}
The large $c$ matter CFT couples to the metric in the standard way and the equations of motion for  the metric,  namely Einstein's equations, are actually the equations obeyed by the dilaton,
\EQ{
\nabla_\alpha\nabla_\beta\phi+R(\phi-\Delta\phi)g_{\alpha\beta}=8\pi G_NT_{\alpha\beta}\ ,
}
or in component form
\EQ{
e^{2\rho}\partial_{w}\big(e^{-2\rho}\partial_{w}\phi)&=8\pi G_N T_{ww}\ ,\\[5pt]
e^{2\rho}\partial_{\bar w}\big(e^{-2\rho}\partial_{\bar w}\phi)&=8\pi G_N T_{\bar w\bar w}\ ,\\[5pt]
-\partial_{w}\partial_{\bar w}\phi+\frac12e^{2\rho}\phi&=8\pi G_N T_{w\bar w}\ .
\label{yfr}
}
This illustrates one of the simplifying features of JT gravity; namely, the CFT stress tensor sources the dilaton but does not couple to it directly. Since we work in the semiclassical approximation, on the right hand side of Einstein's equations are  the expectation values of components of the stress tensor. In particular, the mixed component $T_{w\bar w}=0$ in the semiclassical limit. 

The first of the three equations \eqref{yfr} leads to a condition on the location of the QES. When the derivative hits the conformal factor, there is a simple pole at each of the branch points $w=a_j$, with strength proportional to the derivative of the dilaton evaluated at the branch point. The expectation value of the CFT stress tensor has matching poles, as dictated by  conformal Ward identities  \cite{DiFrancesco:1997nk, Goto:2020wnk} following from the OPEs of the stress tensor with primary fields,  in this case the  twist fields.  The residues of the simple poles are proportional to the derivatives of the CFT R\'enyi entropy with respect to $(a_j,\bar a_j)$:
%\be
%\left(\frac1n-1\right)\partial_w\phi\big|_{w=a_j}
%\ee
 equating the residues gives the QES condition quoted in the introduction \eqref{wii}. Importantly, putative double poles at the locations of the branch point twist fields are cancelled by  the Weyl shifts on top of the flat space stress tensor due to  the nontrivial conformal factor of the metric on the base.

\subsection{Boundary dynamics}
The action of JT gravity reduces purely to a boundary term and all non-trivial dynamics is determined by the boundary map 
\be
w(\tau)=e^{i\theta(\tau)}\,,
\ee
 where $\tau$ is the (Euclidean) bath time. This boundary effective action is related to the ADM mass of the geometry given by the Schwarzian of the boundary map. In the context of the replica geometry, the relevant map is the one in the covering space, namely $W(\tau)=W(w(\tau))$:\footnote{The entropy is related to action of the gravity and CFT via, $(1-n)S_n=\log (Z_n/Z_1)$, where the partition function $\log Z_n=-I^{(n)}_\text{JT}-I^{(n)}_\text{CFT}$.}
\EQ{
I_\text{JT}^{(n)}[w]&=-NS_0-\frac{\phi_r}{8\pi G_N}\int_{\partial\widetilde{\cal M}_n} d\tau\,\{W,\tau\}\\[5pt] 
&=-NS_0-\frac{n\phi_r}{8\pi G_N}\int_{|w|=1} d\tau\,\Big(\{w,\tau\}+2{\EuScript T}(w)(\partial_\tau w)^2\Big)\ .
\label{rff}
}
The final form of the action is defined on the boundary of the base, the circle $|w|=1$, and this accounts for the extra factor of $n$. We have defined the quantity 
\be
{\EuScript T}(w)=\frac12\{W,w\}
\ee
which is a meromorphic function of $w$, and which we also recognize as the Liouville stress tensor 
\be
{\EuScript T}(w)=-(\partial_w \rho)^2 + \partial_w^2\rho. \label{liouville}
\ee
When the gravitational sector has been reduced to the dynamics of the boundary map $w(\tau)$ the bath CFT stress tensor then sources the boundary map. The  equation of motion for the boundary map was derived in \cite{Engelsoy:2016xyb} and also in the pioneering replica wormhole paper \cite{Almheiri:2019qdq} in a different way. Here, we summarize the original. The boundary map determines a coordinate transformation between the near boundary AdS Kruskal-Szekeres coordinates $(w,\bar w)$ and the Euclidean bath coordinates $(y=\sigma+i\tau,\bar y=\sigma-i\tau)$,
\EQ{
w=w(-iy)\ ,\qquad \bar w=\bar w(i\bar y)\ .
}
Let us write the gravitational part of the metric in terms of the bath coordinates as
\EQ{
ds^2=e^{2\tilde\rho}\,|dy|^2\ .
}
Then as one approaches boundary at $\sigma=-\epsilon$, for small $\epsilon$, where the AdS part of the geometry is glued to the Minkowski part, the conformal factor behaves as 
\EQ{
e^{2\tilde\rho}=\frac1{4\epsilon^2}-\frac16\{W,\tau\}+{\mathscr O}(\epsilon^2)\,,
}
where $W=W(w(\tau))$. The fact that the right-hand side of the third equation in \eqref{yfr} vanishes, on account of $T_{w\bar w}=0$, requires the dilaton to have the asymptotic behaviour
\EQ{
\phi=\frac{\phi_r}\epsilon+\phi_0+\frac{2\epsilon\phi_r}3\{W,\tau\}+{\mathscr O}(\epsilon^2)\,.
}
If we substitute this asymptotic form into the first two equations \eqref{yfr} and perform a diffeomorphism to $(y,\bar y)$ and then take the difference, this gives an expression that has a non-singular limit as $\epsilon\to0$ and is the equation we are after, the equation of motion of the boundary map $w(\tau)$ sourced by the CFT:
\EQ{
\frac{\phi_r}{8\pi G_N}\frac d{d\tau}\Big[\{w,\tau\}+2{\EuScript T}(w)\Big(\frac{dw}{d\tau}\Big)^2\Big]=i\big(T_{yy}(i\tau)-T_{\bar y\bar y}(-i\tau)\big)\ .
\label{ebe}
}
On the right, we have the expectation values of the CFT stress tensor components in the bath coordinates $(y,\bar y)$ evaluated at the boundary at $y=i\tau$. In Lorentzian signature, this equation has the interpretation as an energy conservation equation: the left-hand side is the rate of change of the gravitational energy while the right-hand side is the inward energy flux of the matter CFT across the boundary.

\section{One QES example}\label{s2}

The simplest scenario is when there is one QES in the context of the eternal black hole \cite{Almheiri:2019qdq,Goto:2020wnk} as illustrated in figure \ref{fig5}. Whilst this is not directly relevant to  deriving the Page curve for the radiation it is revealing enough to provide answers to the questions that we posed in the introduction in the simplest of settings and so is worth analysing in detail.
\begin{figure}[ht]
\begin{center}
\begin{tikzpicture} [scale=0.6,every node/.style={scale=0.8}]
\draw[yellow!20,fill=yellow!20] (-4,-4) rectangle (4,4);
\draw[thick,fill=pink!20] (0,0) circle (2.5cm);
\filldraw[black] (1,0.7) circle (2pt);
\filldraw[black] (2.8,1.8) circle (2pt);
\draw[thick] (1,0.7) to[out=-170,in=10] (-4,-1);
\draw[thick] (4,1.6) to[out=-170,in=10] (2.8,1.8);
\node at (1,0.3) {$a$};
\node at (2.8,1.3) {$b$};
\begin{scope}[xshift=12cm]
\filldraw[pink!20] (-3,-4) rectangle (3,4);
\filldraw[yellow!20] (6,0) -- (3,3) -- (3,-3) -- (6,0);
\filldraw[yellow!20] (-6,0) -- (-3,3) -- (-3,-3) -- (-6,0);
\draw[-] (-3,-3) -- (-6,0) -- (-3,3);
\draw[-] (3,-3) -- (6,0) -- (3,3);
\draw[-] (-3,-3) -- (3,3);
\draw[-] (-3,3) -- (3,-3);
\draw[-] (3,-4) -- (3,4);
\draw[-] (-3,-4) -- (-3,4);
\filldraw[black] (3.6,1) circle (2pt);
\filldraw[black] (1.6,0.8) circle (2pt);
\draw[thick] (3.6,1) to[out=-10,in=150] (6,0);
\draw[thick] (-6,0) to[out=20,in=-170] (1.6,0.8);
\end{scope}
\end{tikzpicture}
\caption{\footnotesize The set up in (left) Euclidean (right) Lorenztian signature with one QES in the context of the eternal black hole. The AdS region is shaded pink and the bath in yellow. There is a point in the right Euclidean/Minkowski bath and one QES as shown ultimately outside the horizon of the right black hole in the Lorentzian picture.}
\label{fig5} 
\end{center}
\end{figure}
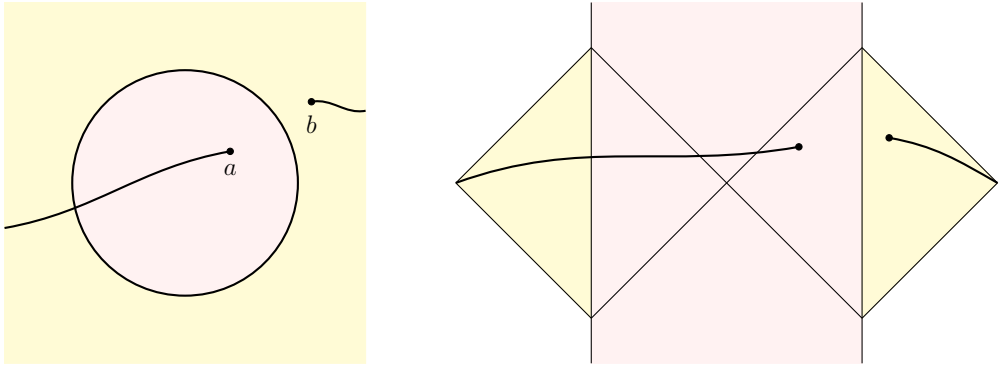

There is one point in the right bath with coordinates $(e^y,e^{\bar y})=(b,\bar b)$. For simplicity, we also re-scale $y\to \beta y/2\pi$, so that in Euclidean signature $\tau$ has period $2\pi$. 

With one branch point, the unformization problem is straightforward and the inverse uniformisation map takes the form
\EQ{
W=\Big(\frac{w-a}{1-\bar aw}\Big)^{1/n}\ .
\label{lak}
}
This is, as expected, multi-valued. The fundamental domain ${\cal D}$ is the whole of the unit disc $|W|=1$. In this case there are no circles as in figure \ref{fig2} to cut out. The boundary $|W|=1$ covers the boundary of the base $|w|=1$ $n$ times. 

\subsection{Off-shell gravity action}
The action for JT gravity evaluated on the replica geometry consists of two parts, 
\EQ{
&I_{\rm JT}^{(n)}[w; a, \bar a]=
-NS_0\\ & -\frac{n\phi_r}{4\beta G_N}\Big(\int_{|w|=1}d\tau\{w,\tau\}+\frac{(n^2-1)}{2n^2}(1-|a|^2)^2\oint_{|w|=1} dw\,\frac{\dot w}{(w-a)^2(1-\bar aw)^2}\Big)\,
\label{ojn1}
}
where, in the above we denote
\EQ{
\dot w(w)=\frac{dw}{d\tau}\Big|_{\tau=\tau(w)}\ ,
}
i.e.~$\dot w$ as a function of $w$.  Off-shell, the Schwarzian term is independent of the branch point location which is moved around by M\"obius transformations on $w$.  The off-shell dependence on the QES location is captured by the second term, provided we only consider boundary maps modulo ${\rm SL}(2,{\mathbb R})$ variations, keeping the branch point fixed.  We have expressed the second term as a contour integral over $w$.\footnote{This expression is re-scaled by $2\pi/\beta$ to reflect the fact that $\tau$ has been re-scaled by $\beta/2\pi$ in order to make the thermal circle have length $2\pi$.}
%\EQ{
%I_\text{JT}=-\frac{(n^2-1)\phi_r(1-|a|^2)^2}{8n\beta G_N}\oint_{|w|=1} dw\,\frac{\dot w}{(w-a)^2(1-\bar aw)^2}\ ,
%\label{ojn}
%}
The purpose of writing the integral in this way is that we can relate it directly to the dilaton. 
\subsection{The dilaton from boundary dynamics}
On-shell the dilaton satisfies the wave equation \eqref{yfr} sourced by the CFT stress tensor. However, we are after an expression for the dilaton in terms of the boundary curve $w(\tau)$ rather than the CFT stress tensor. The boundary curve responds 
to the energy flux and so we can trade in one for the other.
For the un-replicated problem $n=1$ where the metric is just the standard one $ds^2=|dw|^2/(1-|w|^2)^2$, this can be done by using a Green's function method as described in \cite{Goto:2020wnk} and reviewed in appendix \ref{app:dilaton}. 

In the context of the $n>1$ problem, we simply notice that the metric takes this form in the covering space \eqref{zaz} and so the same method applies. In particular, the stress tensor component $T_{WW}$, like $T_{ww}$ in the $n=1$ problem, is analytic in the disc, here in the covering space $|W|\leq1$.\footnote{This follows because in flat space the CFT stress tensor has double poles at the positions of the twist operators. However, these are cancelled by the Weyl transformation to the physical metric in $(w,\bar w)$ coordinates and then a diffeomorphism to the covering space  \cite{Goto:2020wnk}.}
In addition, a key ingredient in the derivation is the equation of motion for the boundary curve and again we remark that in our case \eqref{ebe} can be written in terms of $W(w)$ in a way that is identical to the $n=1$ equation with $w\to W$. We have no need to repeat the steps here, but simply quote the result,
\EQ
{
\phi( W,\bar W)=-\frac{\phi_r(1-| W|^2)^2}{\beta}\oint_{|\tilde W|=1}d\tilde W\,\frac{\dot { W}(\tilde W)}{( \tilde W-W)^2(1-\bar W\tilde W)^2}\,.
\label{jjw1}}
In the above, $\dot {W}=\partial_w W(w)\dot w(w)$ and then mapped back to the covering space using the uniformisation map $\pi:\ w\to W$. We have
\EQ{
\partial_w W=\frac{(1+\bar a W^n)^2}{n(1-|a|^2) W^{n-1}}\,.\label{dWw}
}
In section \ref{s4}, we show how the  above integral expression for the dilaton generalizes for multiples QES, or branch points.

The integral expression \eqref{jjw1}, together with $\eqref{dWw}$, manifests the dependence of the dilaton on the position of the QES.  As we discuss below,  it can be computed by picking up the residues from the double pole at $\tilde W= W$ and the pole of order $n-1$ at the branch point $\tilde W=0$ or, at the mirror points $\tilde W=1/\bar W$ and $\infty$.

For now,  we only need the expression for the dilaton evaluated at the QES, corresponding to the branch point. In this case the two poles collide and leave a much simpler integral over the boundary of the base instead,
\EQ{
\phi(a,\bar a)\equiv \phi( W=0,\bar W=0)=-\frac{\phi_r(1-|a|^2)^2}{n\beta}\oint_{|w|=1}dw\,\frac{\dot w}{(a-w)^2(1-\bar aw)^2}\ .
\label{jns21}
}
In deriving this expression, one should remember that the boundary in the unformized space $|W|=1$ covers the boundary in the base $n$ times. 

Now, we notice that the integral expression for the dilaton evaluated at the QES \eqref{jns21} is proportional to the second term in the gravitational action \eqref{ojn1}, so that
\EQ{
I_\text{JT}^{(n)}[w; a,\bar a]=-NS_0-\frac{n\phi_r}{4\beta G_N}\int_{|w|=1}d\tau\{w,\tau\}+\frac{(n^2-1)\phi(a,\bar a)}{8G_N}\ .
\label{pey}
}
It is worth stressing that as an off-shell expression for the semiclassical gravity action on the replica geometry, the explicit dependence on the branch point location is encoded only in the dilaton term. Formally including the  contribution from the CFT bath in the semiclassical limit, we can write full replica partition function as 
\be
-\log Z_n =  I_{\rm JT}^{(n)}[w, a,\bar a] -\log Z_n^{\rm CFT}
\ee
The CFT partition function depends on the location of the QES in the AdS$_2$ region,   and implicitly on the boundary map $w(\tau)$ in a way which is determined by the solution to the welding problem, which we will flesh out further below.  
\subsection{Generalised modular entropy}
Varying the off-shell action with respect to the location of the branch point yields the QES condition
\be
\partial_a\left(\frac{\phi(a,\bar a)}{4G_N} +\frac{2}{n+1} S_n^{\rm CFT}\right)=0\,.
\ee
The modular entropy, defined as
\be
\widetilde S_n^{\rm CFT} = - n^2 \frac{\partial}{\partial n} (n^{-1}\log Z_n^{\rm CFT}),
\ee
can be written in terms of the R\'enyi entropy in the case of free fermions, and more generally for one interval problem, as 
\be
\widetilde S_n^\text{CFT}=\frac2{n+1} S_n^{\rm CFT}\,.
\ee
The QES condition above makes it clear that the appropriate generalized modular entropy is
\EQ{
\widetilde S_n^\text{gen}(a,\bar a)=\frac{\phi(a,\bar a)}{4G_N}+\widetilde S_n^\text{CFT}(a,\bar a)\ ,
\label{des}
}
in complete agreement with Dong's holographic R\'enyi entropy, which generalizes  the RT formula \cite{Dong:2016fnf} once we identify the  dilaton as playing the r\^ole of the area.
\subsection{Consistency with equation of motion}
We can reach the same conclusion by  examining condition \eqref{wii} following from the JT equations of motion and the CFT Ward identity relating the simple pole residue in the stress tensor (with twist field insertions) to the CFT R\'enyi entropy. 

We use the integral expression in the cover \eqref{jjw1} to compute the derivative 
$\partial_w\phi(w,\bar w)$ and then evaluate it at $w=a$. The calculation is easiest to perform in covering space coordinates, picking up the residues at the poles of the integrand. This is straightforward provided $\dot w(w)$ and its analytic structure is known. In general, however, $\dot w(w)$ is known only formally via a  Laurent expansion, 
\be
\dot w (w) = \dot w_+(w) + \dot w_-(w) \label{spl}
\ee
where $\dot w_+$ and $\dot w_-$ are analytic inside and outside the unit disc respectively,
\be
\dot w_+(w) = i w + \sum_{m>1} p_{m}^+\, w^m \,,\qquad \dot w_-(w)=\sum_{m>1} p_{m}^- \, w^{-m+1} \,.\label{laurent}
\ee
 Viewed as functions of the covering space coordinate $W$, $\dot w_+$ has poles outside the unit disc, whist the singularities of $\dot w_-$ lie inside the unit disc. The integrand in \eqref{jjw1}   has double poles inside the unit disc, at $\tilde W = W$, and outside the disc at $\tilde W = 1/\bar W$, assuming $|W|<1$.  There is also a pole of order $n-1$ at $\tilde W=0$ from $\partial_w W$ expressed as a function of $\tilde W$.

We then compute the integral \eqref{jjw1} by splitting it in two parts. We evaluate the piece dependent on $\dot w_+$ by picking up the residues {\em inside} the unit disc, whilst the second integral involving $\dot w_-$ is computed by picking up the residues  from poles {\em outside} the unit disc. The dilaton and its derivative at the branch point are both well defined and given as
\EQ{
\phi(a,\bar a)=-\frac{2\pi i\phi_r}{n\beta}\left(\partial_w\dot w_+(a) -\partial_w\dot w_-(\bar a ^{-1})+ \frac{2 \bar a}{1-|a|^2}\left(\dot w_+ (a) +\dot w _-(\bar a ^{-1})\right)\right)\label{diln1}
}
and 
\EQ{
\partial_w\phi(a,\bar a) = -\frac{n+1}{n}\cdot\frac{2\pi i\phi_r}{\beta }&\Big(\frac{\bar a^2}{(1-|a|^2)^2} (\dot w_+(a)+\dot w_-(\bar a^{-1}) \\[5pt]
&+\frac{\bar a\,\partial_w\dot w_+(a)}{1-|a|^2}  +\frac12\partial_w^2\dot w_+(a)\Big)\,.
\label{diln2}
}
It follows immediately that 
\EQ{
\partial_w\phi(a,\bar a)=\frac{n+1}2\partial_a\phi(a,\bar a)\,.
}
Given how carefully we have to account for the singularity structure of $\dot w (w)$, the match is quite nontrivial. Thus eq.\eqref{wii} which is a consequence of the CFT stress tensor Ward identity, becomes a QES equation for the generalised modular entropy \eqref{des},
\be
\partial_a \tilde S^{\rm gen}(a,\bar a)=0\,,
\ee
valid for any $n$. This shows precisely how the puzzle posed in the introduction is solved in detail. The dilaton $\phi(w,\bar w)$ has implicit dependence on the position of QES $(a,\bar a)$ as well as explicit dependence when evaluated at the QES. Hence, there is no contradiction between the Ward identity equation \eqref{wii} and the variational equation \eqref{wii2} inspired by Dong's R\'enyi generalization of the RT formula. We have thus shown that the notions of QESs and fine grained entropies admit natural generalisations for R\'enyi  entropies with arbitrary $n$. 
\subsection{Welding problem and boundary map}
In order to write down the equation of motion for the boundary map \eqref{ebe}, we need the expectation value of the  stress tensor of the CFT, which for simplicity we assume is a set of $c$ massless fermions. Therefore we need to  know what is the appropriate quantum state of the CFT.  

Na\"\i vely, one would think that the thermal state in the bath coordinates $(y,\bar y)$ along with the insertions of the twist operators at branch points, would be the appropriate choice. However, with the gluing provided by the boundary map $w(\tau)=e^{i\theta(\tau)}$ this would generally not extend to a holomorphic map in the AdS region. So one has to choose the state to be the vacuum state in some more general coordinates $(z,\bar z)$ that are related holomorphically to $(y,\bar y)$ in the bath region, 
\be
z=F(e^y)\,, \qquad {\rm Re}(y) \geq 0 \,,
\ee 
and holomorphically to $(w,\bar w)$ in the the AdS region, 
\be
z=G(\omega)\,,\qquad |w| \leq 1\,.
\ee
 Finding this coordinate frame is the conformal welding problem described in detail in \cite{Almheiri:2019qdq}.  The CFT stress tensor along the boundary, in the presence of twist and anti-twist operators at $z=G(a)$ and $z=F(b)$, can then be expressed via the Schwarzian derivative
\EQ{
T_{yy}=-\frac c{24\pi}\Big\{\Big(\frac{F(e^y)-F(b)}{F(e^y)-G(a)}\Big)^{1/n},y\Big\}\ ,
}
with a similar expression for the anti-holomorphic component. This is to be inserted on the right-hand side of the equation of motion \eqref{ebe}. 

The equation of motion depends on the conformal welding problem in a non-trivial way and it becomes impossible to find the solution $w(\tau)$ in general. Some form of approximation is needed in order to make progress. One way to proceed is to work in perturbation theory in $n-1$ with a view to taking the $n\to1$ limit in order to compute the von Neumann entropy. This provides a tractable limit and leads to the familiar generalized (von Neumann) entropy and QES formula but it is not a regime open to our analysis where we want to study the problem for integer $n$. 

Fortunately there is another approximation that is open to us and which works for any $n$; namely, the high temperature limit where the dimensionless combination
\EQ{
\kappa\equiv \frac{c\beta G_N}{6\pi\phi_r}\ll1\, \label{kappa}
}
$\kappa$ controls the strength of the backreaction due to the energy flux in eq.\eqref{ebe}.
This is the limit analysed in \cite{Almheiri:2019qdq} and corresponds to a large Bekenstein-Hawking entropy $S_\text{BH}(\beta)\gg c$ and physically to when the absorption and emission of CFT modes by the black hole is very slow, precisely the quasi-static regime where Hawking's original calculation is valid.  

In the small $\kappa$ regime, the boundary map is determined in perturbation theory around the the trivial map. To linear order, with $\delta\theta\sim {\mathscr O}(\kappa)$, 
\EQ{
w(\tau)=e^{i\theta(\tau)}=e^{i\tau}\big(1+i\delta\theta(\tau)\big)\ ,\qquad \delta\theta(t)=\sum_{m\neq0,\pm1}c_me^{im\tau}\ ,
\label{bed}
}
which has the requisite periodicity in imaginary time (after the re-scaling by $\beta/2\pi$). Crucially, we omit the modes with $m=0,\pm1$ which are fixed by the $\text{SL}(2,{\mathbb R})$ symmetry that acts as M\"obius transformations on $w$. These are precisely the transformations that move the branch point around. Off-shell, we want to keep $w(\tau)$ independent of variations in the location of the QES.

The solution to the welding problem can be expressed in terms of the expansions,
\EQ{
G(w)=w\Big(1-i\sum_{m>1}c_mw^m\Big)\ ,\qquad\quad F(e^y)=e^{y}\Big(1+i\sum_{m<-1}c_me^{my}\Big)\,
\label{hcc}
}
valid in the domains $|w|<1$ and ${{\rm Re} (y) > 0}$, respectively.
%The CFT is in the vacuum state for both infalling and outgoing modes in the $(z,\bar z)$ coordinate frame and so the stress tensor can then be expressed via the Schwarzian derivative
%\EQ{
%T_{yy}=-\frac c{24\pi}\big\{\Big(\frac{F(e^y)-F(b)}{F(e^y)-G(a)}\Big)^{1/n},y\big\}\ ,
%}
%with a similar expression for the anti-holomorphic component. We can insert these expressions into the right-hand side of the equation of motion \eqref{ebe} which is then of order $\kappa$. 

As will become apparent once we solve for the position of the QES, $a$ and $\bar a$ are order $\kappa$. With this in mind, we first solve the equation of motion \eqref{ebe} at order $\kappa$ for all the Fourier modes of $\delta\theta(\tau)$ except $m=0,\pm1$, thus keeping the QES location off-shell for now, so
%takes the form
%\EQ{
%\frac{\phi_r}{8\pi G_N}\frac d{d\tau}\Big[\{w,\tau\}+\frac{(n^2-1)(1-|a|^2)\dot w^2}{n^2(w-a)^2(1-\bar aw)^2}\Big]
%\ .
%\label{fly}
%}
%The second term here, is not obviously of order $\kappa$. Note that it is order $n-1$ and so when one is computing the von Neumann entropy, this term, like the matter term on the right-hand side is small and perturbation theory in $n-1$ can be applied as in \eqref{bed}. However, here are working with $n$ integer and instead we are leveraging the high temperature approximation, i.e.~small $\kappa$.  As will only become apparent once we solve for the position of the QES and find that $a$ and $\bar a$ are order $\kappa$, the second term is actually of order $\kappa^2$. 
%Assuming this, at linear order in $\kappa$ we have\footnote{Note that we do not impose the $m=0,\pm1$ equations in order to keep the QES off shell for now.}
\EQ{
\delta\theta(\tau)=-\frac{i\kappa(n^2-1)}{2}\sum_{m>1}\frac{m+1}{m^2((mn)^2-1)}\frac{e^{im\tau}}{b^m}+\text{c.c.}
\label{gil}
}
Now we can compute the gravitational action \eqref{ojn1} with the QES off-shell.
% The Lagrangian involves
%\EQ{
%\{\Big(\frac{w-a}{1-\bar aw}\Big)^{1/n},\tau\}=\{w,\tau\}+\frac{(n^2-1)(1-|a|^2)^2\dot w^2}{2n^2(w-a)^2(1-\bar aw)^2}\ .
%}
%Hence for one QES, we have
%\EQ{
%{\EuScript T}(w)=\frac{(n^2-1)(1-|a|^2)^2}{4n^2(w-a)^2(1-\bar aw)^2}\ .
%}
%In particular, the accessory parameters are known exactly in this case,
%\EQ{
%c^\pm=\mp\frac{3}{8(a-1/\bar a)}\ .
%\label{lob}
%}
The term involving the Schwarzian $\{w,\tau\}$ vanishes to leading order in $\kappa$ when integrated around the boundary and so we focus on the second term,
%involving ${\EuScript T}(w)$. For the moment we don't make any particular assumption %about the boundary map and we can express the integral of the second term as 
expressed as a contour integral over $w$.
%\footnote{This expression is re-scaled by $2\pi/\beta$ to reflect the fact that $\tau$ has been re-scaled by $\beta/2\pi$ in order to make the thermal circle have length $2\pi$.}
%\EQ{
%I_\text{JT}=-\frac{(n^2-1)\phi_r(1-|a|^2)^2}{8n\beta G_N}\oint_{|w|=1} dw\,\frac{\dot w}{(w-a)^2(1-\bar aw)^2}\ ,
%\label{ojn}
%}
%where, in the above we denote
%\EQ{
%\dot w(w)=\frac{dw}{d\tau}\Big|_{\tau=\tau(w)}\ ,
%}
%i.e.~as a function of $w$. 
Following from \eqref{gil},  to linear order in $\kappa$ we can determine the pieces of $\dot w_\pm(w)$ which are analytic inside and outside the unit disc:
\EQ{\label{analytic}
\dot w_+(w)&=i\omega+\frac{i\kappa(n^2-1)}2\sum_{m>1}\frac{m+1}{m((mn)^2-1)}\frac{w^{m+1}}{b^m}\ ,\\[5pt]
\dot w_-(w)&=-\frac{i\kappa(n^2-1)}2\sum_{m>1}\frac{m+1}{m((mn)^2-1)}\frac{w^{-m+1}}{\bar b^m}\ .
}
Note that the leading term $iw$ can be placed in either of $\dot w_\pm$. 
Then the integral \eqref{ojn1} can be computed by by picking up the residue at $w=a$ for the $\dot w_+$ component and $w=1/\bar a$ for the $\dot w_-$ component to yield
\EQ{
\phi(a,\bar a)&=\frac{2\pi\phi_r}{n\beta}\frac{1+|a|^2}{1-|a|^2}\\[5pt] &-\frac{\pi\kappa\phi_r(n^2-1)}{n\beta(1-|a|^2)} \sum_{m>1}\frac{(m+1)(m(1-|a|^2)+1+|a|^2)}{m((mn)^2-1)}\Big(\frac{a^{m}}{b^m}+\frac{\bar a^{m}}{\bar b^m}\Big)\ .
\label{jns3}
}
In the $n\to1$ limit, the first term is the well-known expression for the dilaton in the Euclidean eternal black hole. 

\subsection{QES at next-to-leading order}

Since $a$ is effectively order $\kappa$, the next-to-leading terms in \eqref{jns3} are those with $m=2$. Keeping these terms gives
\EQ{
\phi(a,\bar a)=\frac{2\pi\phi_r}{n\beta}\Big(1+2|a|^2+\frac{9\kappa(n^2-1)}{4(4n^2-1)}\frac{(a^2\bar b^2+\bar a^2b^2)}{|b|^4}+\cdots\Big)\ .
\label{jns4}
}
To this order, the solution of the conformal welding problem involves
\EQ{
\label{weld}
G(w)&=w-\frac{3\kappa(n^2-1)}{8(4n^2-1)b^2}w^3+\cdots\ ,\\[5pt] F(e^y)&=e^y-\frac{3\kappa(n^2-1)}{8(4n^2-1)\bar b^2}e^{-y}+\cdots\ .
}
These functions appear in the CFT modular entropy,
\EQ{
\widetilde S^\text{CFT}_n(a,\bar a)=\frac{c}{6n}\log\frac{|F(b)-G(a)|^2}{(1-|a|^2)|G'(a)F'(b)b|}\ .
}
In order to extremize the generalized entropy we will need the effective expansion in powers of $\kappa$
\EQ{
\partial_a\widetilde S_n^\text{CFT}(a,\bar a)=\frac{c}{6n}\Big(-\frac1b+\bar a-\frac{a}{b^2}-\frac{3\kappa(n^2-1)}{8(4n^2-1)|b|^4b}+\cdots\Big)\ .
}
Extremizing the generalized modular entropy \eqref{des} gives the position of the QES to the next-to-leading order
\EQ{
a=\frac\kappa{\bar b}+\frac{\kappa^2(1-|b|^2)}{|b|^2\bar b}+\frac{3\kappa^2(n^2-1)}{8(4n^2-1)}\frac{1-6|b|^2}{|b|^4\bar b}+\cdots\ .
}
The modular entropy of the saddle to the order we are working is equal to
\EQ{
\widetilde S^\text{gen}_n=\frac{2\pi\phi_r}{n\beta}\Big(1+\log|b|-\frac{2\kappa}{|b|^2}-\frac{9\kappa(n^2-1)}{8(4n^2-1)|b|^4}+\cdots\Big)\ .\label{sgen}
}
As anticipated $a$ is of order $\kappa$ and the next-to-leading order terms are $n$ dependent. We have computed this in Euclidean signature but we can analytically continue to Lorentzian signature to capture the real-time entropy dynamics. In particular, the QES is at $a^+=-\kappa/b^-$ and $a^-=-\kappa/b^+$ and since $b^\pm\gtrless0$, means that the QES will be outside the horizon as shown in figure \ref{fig5}.

\subsection{Dilaton at arbitrary point}

Before we leave the single QES analysis, it is interesting also to evaluate the dilaton at an aribtrary point in Lorentzian signature. When $n=1$, one can confirm that the expression is simply
\EQ{
\phi(w^\pm)_{n=1}=\frac{2\pi\phi_r}\beta\frac{1-w^+w^-}{1+w^+w^-}\ ,
}
independent of the position of the QES as expected since the back-reaction from the QES vanishes.
For general $n$, the general expression is complicated but near the QES, $w^\pm\sim a^\pm$, we have, for $n>2$,\footnote{The $n=2$ expression is slightly more complicated.}
\EQ{\label{dilgen}
\phi(w^\pm)=\frac{2\pi\phi_r}{n\beta}\Big\{\frac{1-a^+a^-}{1+a^+a^-}+\frac{2[(w^+-a^+)(a^--w^-)]^{\frac1n}}{(1+a^+a^-)^{1+\frac2n}}+\cdots\Big\}\ .
}
Of course when evaluated at the QES we return to the expression \eqref{jns3} continued to Lorentzian signature. Notice that this expression is real for points space-like separated from the QES ($w^+\gtrless a^+$ and $a^-\gtrless w^-$) but complex for points time-like separated from the QES. This is in line with the general analysis of \cite{Colin-Ellerin:2021jev} of real-time replica wormholes which found that the metric, or for JT gravity the dilaton, will be complex in the causal past of the splitting surface, the QES. In fact, if we evaluate the dilaton with the leading order position for the QES $a^\pm=0$, we find
\EQ{
\phi(w^\pm;a^\pm=0)=\frac{2\pi\phi_r}{n\beta}\frac{1-(-w^+w^-)^{1/n}}{1+(-w^+w^-)^{1/n}}\ ,
}
which is equivalent to the expression for the dilaton in \cite{Colin-Ellerin:2021jev}.

\section{Replica wormhole and uniformisation}\label{s3}

In this section, we describe the geometrical set up required to construct the replica wormhole when there is more than one QES. The mathematical setting is the theory of Fuchsian uniformisation described in a physics context in \cite{Hulik:2016ifr,Hadasz:2005gk,Hadasz:2006rb} and in the slightly different context of Schottky uniformisation and AdS/CFT for CFTs in $1+1$ dimensions \cite{Faulkner:2013yia}. 

In order to describe the covering geometry $\widetilde{\cal M}_n$, it is useful to define an abstract auxiliary problem in the form of a Fuchsian equation, a second order differential equation on the full $w$ complex plane,
\EQ{
\big(\partial_w^2+{\EuScript T}(w)\big)\Psi=0\ ,
\label{fuc}
}
along with its complex conjugate. The relation to the replica wormhole is expressed through the potential function ${\EuScript T}(w)$ that has already appeared in the gravitational action \eqref{rff} and which is identified with the holomorphic component of the Liouville stress tensor \eqref{liouville}. 

As an analytic function this has double poles at the QES $w=a_j$ on the disc and image points $w=1/\bar a_j$ outside the disc of a specific form
\EQ{
{\EuScript T}(w)=\sum_{j=1}^N\Big(\frac{\varepsilon_n}{(w-a_j)^2}+\frac{\varepsilon_n}{(w-1/\bar a_j)^2}+\frac{c^+_j}{w-a_j}+\frac{c^-_j}{w-1/\bar a_j}\Big)\ ,
\label{der}
}
where, 
\be
\varepsilon_n=\frac{n^2-1}{4n^2}\,.\label{varep}
\ee
The $(c^+_j,c^-_j)$ are the {\it accessory parameters\/} which depend implicitly on the coordinates $(a_j,\bar a_j). $\footnote{The accessory parameters satisfy constraints ensuring the absence of an additional singularity at $w=\infty$. For $N>1$, the total number of undetermined accessory parameters is $2N-3$.}
The dependence of the accessory parameters on the $(a_j,\bar a_j)$ can be uncovered in an  expansion around a certain factorization limit. For the case with $N=1$, which we have studied in detail above, the accessory parameters are completely determined by the requirement that $w=\infty$ is a non-singular point,
\be
N=1:\qquad c^\pm= \pm  \varepsilon_n \frac{2\bar a}{1-|a|^2}\,.\label{accbasic}
\ee
The fact that branch points occur in pairs $(a_j,1/\overline a_j)$, one inside and one outside the unit circle, is needed so that the conformal factor has the required behaviour \eqref{dew}  near the boundary as $|w|\to1$. On the boundary, the stress tensor satisfies the reality conditions 
\EQ{
w^2{\EuScript T}(w)\in{\mathbb R}\ ,\qquad {\EuScript T}(w)=\frac1{w^4}\,\overline{{\EuScript T}(1/\bar w)}\ .
\label{exx}
}
Let us denote a pair of two linearly independent solutions of the Fuchsian equation \eqref{fuc} with constant Wronskian, the fundamental system, $\Psi=(\psi_1,\psi_2)$. In order to satisfy the reality conditions \eqref{exx} we must impose the following reality conditions on the fundamental system
\EQ{
w\,\overline{\psi_1(1/\bar w)}=\psi_2(w)\ ,\qquad w\,\overline{\psi_2(1/\bar w)}=\psi_1(w)\ .
\label{rel}
}
Given a fundamental system that does not satisfy these conditions it is always possible to find an $\text{SL}(2,{\mathbb C})$ transformation $\Psi\to V\cdot\Psi$ so that the transformed system does satisfy \eqref{rel}. Once \eqref{rel} is satisfied it is fixed by the subgroup $\text{SU}(1,1)$ which will play a prominent r\^ole as we proceed.

This fundamental system has non-trivial monodromies $(M_j,M_{\bar j})\in\text{SU}(1,1)$ around each of the branch points $(a_j,1/\bar a_j)$ on the complex $w$ plane. Note that $\text{SU}(1,1)$ is precisely the subgroup of M\"obius transformations that preserves the unit disc. The monodromies satisfy $M_j^n=M_{\bar j}^n=I$ to reflect the ${\mathbb Z}_n$ symmetry and the coefficients of the double poles in \eqref{der} means that they must lie in the conjugacy class of the element 
\EQ{
\Lambda\equiv\text{diag}(-e^{\pi i/n},-e^{-i\pi/n})\ .
}
What makes the problem non-trivial is that the monodromies in general do not commute $[M_i,M_j]\neq0$. However, the monodromies around each pairs of points $(a_j,1/\bar a_j)$ are trivial 
\EQ{
M_{\bar j}M_j=I\ .
}
This means that the mirror points $1/\bar a_j$ have monodromies in the conjugacy class of $\Lambda^{-1}$.

The connection to the replica wormhole is that the solution to Liouville's equation \eqref{ldr}, the conformal factor of the metric, is given in terms of the fundamental system via
\EQ{
e^{2\rho}=\big(|\psi_1|^2-|\psi_2|^2\big)^{-2}\ .
\label{ab1}
}
This is invariant under the mondromies and so $\rho$ is well defined on the base. 

\subsection{Fuchsian uniformisation}

At this point, we can give a uniformisation perspective to the construction. The idea is to realize the Riemann surface as the unit disc $|W|=1$ modulo the action the Fuchsian group which is a discrete subgroup $\Sigma\subset\text{SU}(1,1)$.\footnote{Our discussion will not address some of the subtleties of this construction.} These transformations take the form
\EQ{
W\longrightarrow\frac{uW+v}{\bar vW+\bar u}\ ,\qquad |u|^2-|v|^2=1\ .
\label{rmm}
}
One can think of the generators of $\Sigma$ as being associated to a set of $g$ homology cycles that identify pairs of circles $(C_a,\tilde C_a)$ in the complex $W$ plane to create a genus $g$ surface. The unit disc with the circles cut out describes a  fundamental domain ${\cal D}$ for the cover. The boundary $\partial{\cal D}$ consists of a subset of the boundary of the disc $|W|=1$ along with a series of arcs, the boundaries of the circles $(C_a,\tilde C_a)$ that lie in unit disc. See figure \ref{fig2} for an example with $n=3$ and two QES. In this case there are 3 pairs of circles. 

The uniformation map is defined as $\pi:\,{\cal D}\to {\cal M}_n$, i.e.~$W\mapsto w$. We will often consider the inverse map $W(w)$ which is $n$-fold valued determined by the fundamental system
\EQ{
W(w)=\psi_1(w)/\psi_2(w)\ .
\label{ab2}
}
More generally we can exploit the $\text{SU}(1,1)$ symmetry \eqref{rmm} to define a different coordinate in cover $W=(u\psi_1+v\psi_2)/(\bar v\psi_1+\bar u\psi_2)$. On account of \eqref{rel}, the coordinate $W$ in \eqref{ab2} or its $\text{SU}(1,1)$ transformation satisfies the reality condition $\overline{W(1/\bar w)}=1/W(w)$ and that ensures that $W(w)$ maps the unit circle to the unit circle.

The stress tensor \eqref{liouville}, the potential of the Fuchsian equation, is identified with the Schwarzian derivative of the map
\EQ{
{\EuScript T}(w)=\frac12\{W(w),w\}\ ,
\label{sch}
}
which is also invariant under M\"obius transformations and so is valued on the base. This is the function that appears in the gravitational action \eqref{rff}.

Let us now elaborate on some of the details of the Fuchsian uniformisation. By using the freedom to perform $\text{SU}(1,1)$ transformations, we can diagonalize the monodromy of one of the branch points, say $w=a_j$, $M_j=\Lambda$. This has an associated choice of uniformisation coordinate $W_j$ for which the image of the branch point $a_j$ is at $W_j(a_j)=0$ and that of $1/\bar a_j$ at $W_j(1/\bar a_j)=\infty$. The local behaviour around the branch point is
\EQ{
W_j=\lambda_j(w-a_j)^{1/n}\Big(1+\frac{c^+_j}{2n\varepsilon_n}(w-a_j)+{\mathscr O}(w-a_j)^2\Big)\ .
\label{zee}
}
Note the appearance here of the accessory parameter. There is a similar expansion around $w=1/\bar a_j$.

With this choice of coordinate for the cover, the other branch points $w=a_k$ and $1/\bar a_k$, $k\neq j$, have $n$ images related by multiplication by $n^\text{th}$ roots of unity that lie on $\partial{\cal D}$ and are identified by elements of the group $\Sigma$. These are the fixed points of the monodromies $M_j^pM_kM_j^{-p}$, $p=0,1,\ldots,n-1$. If we write the monodromy as 
\EQ{
M_k=U\Lambda U^{-1}\ ,\qquad U=\MAT{u & v\\ \bar  v & \bar u}\in\text{SU}(1,1)\ ,
\label{ger}
}
with $|u|^2-|v|^2=1$, then the $n$ images of $a_k$ are at $W_j(a_k)= e^{2\pi ip/n}v/\bar u$ and of $1/\bar a_k$ at $W_j(1/\bar a_k)= e^{2\pi ip/n}u/\bar v$, $p=0,1,\ldots,n-1$. The coordinates $W_j$ and $W_k$ (i.e.~one for which $W_k(a_k)=0$) are related by the $\text{SU}(1,1)$ transformation 
\EQ{
W_j=\frac{uW_k+v}{\bar vW_k+\bar u}\ .
\label{rsu}
}

There is a natural choice of homology cycles of the Riemann surface. There are a set of $g$ $A$-cycles that we can take as surrounding the images of $(A_k,1/\bar A_k)$, $k\neq j$, on the $p^\text{th}$ copy of the base, $p=0,1,\ldots,n-2$. By construction the monodromy around the $A$-cycles is trivial:
\EQ{
\big(M_j^pM_kM_j^{-p}\big)\big(M_j^pM_{\bar k}M_j^{-p}\big)=I\ .
}
On the other hand, a suitable choice of $B$-cycles have monodromies
\EQ{
L_{k,p}=\big(M_j^pM_kM_j^{-p}\big)M_j\ , \qquad k\neq j\ ,\qquad p\in\{0,1,\ldots,n-2\}\ .
}
These monodromies generate a representation of the Fuchsian group $\Sigma$ and the Riemann surface is then defined by identifying points on boundary $\partial{\cal D}$ with the action of the group. 

As we have previously mentioned there is freedom in specifying the boundary $\partial{\cal D}$. As in \cite{Faulkner:2013yia}, which considered the related Schottky uniformisation problem, there is choice for $\partial{\cal D}$ which respects the $\mathbb Z_n$ symmetry of the replicas. We now describe this choice in more detail. With \eqref{ger} we have
\EQ{
L_{k,p}=\MAT{\gamma|u|^2-|v|^2 & \gamma^p(\gamma^{-1}-1)uv \\ \gamma^{-p}(\gamma-1)\bar u\bar v &  
\gamma^{-1}|u|^2-|v|^2}\ ,
}
where $\gamma=e^{2\pi i/n}$. The pairs of circles $\tilde C_{k,p}=L_{k,p}(C_{k,p})$ are then given by
\EQ{
W_j=-\frac{\gamma^{-1}|u|^2-|v|^2+\mu e^{i\phi}}{\gamma^{-p}(\gamma-1)\bar u\bar v}\ ,\qquad 
\frac{\gamma|u|^2-|v|^2+\mu^{-1} e^{i\tilde\phi}}{\gamma^{-p}(\gamma-1)\bar u\bar v}\ .
}
Here, $\mu$ is parameter that fixes the scales of the circles. The choice $\mu=1$ corresponds to the choice that has the ${\mathbb Z}_n$ symmetry. In that case, the two circles $C_{k,p+1}$ and $\tilde C_{k,p}$ intersect at two points $W_j=\gamma^pv/\bar u$ and $\gamma^pu/\bar v$, precisely the images of the branch points $(a_k,1/\bar a_k)$. We remark that the symmetric choice $\mu=1$ is a non-standard definition of the fundamental domain ${\cal D}$. The same issue was described in the context of Schottky uniformisation in \cite{Faulkner:2013yia}. In our example in figure \ref{fig2}, there are 3 elements of $\Sigma$ needed which identify the coloured circles. One the cycles is shown in blue and note how it is only closed modulo the action of the element of $\Sigma$ on the boundary of the domain. Once the idenifications are made we can visualize the resulting smooth surface as in figure \ref{fig3}.

\section{The case $n=2$ and $N=2$}\label{s2.2}

For the case with $n=2$ and $N=2$, the Riemann surface has genus one and so the map $W(w)$ can be written explicitly in terms of elliptic functions, following a similar analysis to the Schottky uniformisation problem studied in \cite{Faulkner:2013yia}.  It is worth exploring this case in some detail. A pair of independent solutions of the Fuchsian equation \eqref{fuc} take the form
\EQ{
\psi_\pm(w)=\frac1{\sqrt{t'(w)}}e^{\pm t(w)}\ ,
}
where $t(w)$ is an incomplete elliptic integral of the first kind,
\EQ{
t(w)=\alpha\int^w_{-1} \frac{d\tilde w}{\sqrt{\prod_j(\tilde w-w_j)}}\ ,
\label{xox}
}
with $w_j=(1/\bar a_1,a_1,a_2,1/\bar a_2)$. The lower limit of the integral is arbitrary but has been chosen for convenience to be fixed under the transformation $w\to1/\bar w$ that will be important below.\footnote{We can write $t(w)$ explicitly in terms of the incomplete elliptic integral  $F(\varphi,k^2)$ as $t(w)=\frac{i\pi}{2K(x)}\big(F(\varphi(w),k^2)-F(\varphi(-1),k^2)\big)$,  where $k^2= x$ and $\sin^2\varphi(w)= \frac{w_{21}}{w_{42}}\frac{w_4-w}{w-w_1}
%\frac1k\frac{(k+1)w_{43}(w-w_1)+(k-1)w_{13}(w-w_4)}{(k+1)w_{43}(w-w_1)-(k-1)w_{13}(w-w_4)}
$.} The parameter $\alpha$ is fixed by demanding that $W$ has trivial monodromy around either of the two branch cuts along  $(1/\bar a_1,a_1)$, and $(a_2,1/\bar a_2)$. We will refer to this as the $A$-cycle of the homology basis and the requirement is,\footnote{In principle, the right-hand side could be multiplied by any integer but the required solution is the minimal one \cite{Faulkner:2013yia}.}
\EQ{
\alpha \oint_A\frac{d\tilde w}{\sqrt{\prod_j(\tilde w-w_j)}}=2\big(t(a_1)-t(1/\bar a_1))=\pi i\ ,
}
which fixes $\alpha$ as,
\EQ{
\alpha=\frac{\pi\sqrt{w_{31}w_{42}}}{4K(x)}\,, \qquad \qquad K(x)=\int_0^1\frac {d\theta}{\sqrt{1-x\sin^2\theta}}\,.\label{alpha}
}
Here $w_{ij}=w_i-w_j$, and $K(x)$ is the complete elliptic integral, with the conformal cross-ratio $x$ defined as,
\EQ{
x=\frac{w_{21}w_{43}}{w_{31} w_{42}}=\frac{(|a_1|^2-1)(|a_2|^2-1)}{|a_1\bar a_2-1|^2}\ .
\label{fus}
} 
Notice that the cross-ratio is  real.
The monodromy around the   $B$-cycle of the homology basis that surrounds the interval $(a_1,a_2)$  is $\psi_\pm\to q^{\pm1}\psi_\pm$ where
\EQ{
\begin{pmatrix}
\psi_+\\ \psi_-
\end{pmatrix}
\to \begin{pmatrix} q\,\psi_+\\ \frac1q\, \psi_- \end{pmatrix}
\,,\qquad q=e^{-\pi K(1-x)/K(x)}\ ,
\label{defq}
}
where we used the fact that $2\big(t(a_2)-t(a_1)\big)=\pi K(1-x)/K(x)$. In the covering space this is the monodromy around the cycle that leads to the identification of the circles in figure \ref{fig2}.

Now we turn to the reality of the fundamental system $\psi_\pm$. One can easily verify that
\EQ{
\overline{t(1/\bar w)}=t(w)\,,
}
and so $w\,\overline{\psi_\pm(1/\bar w)}=\psi_\pm(w)$, in contrast to \eqref{rel}. However, it is straightforward to find the $\text{SL}(2,{\mathbb C})$ transformation that produces a fundamental system that {\it does\/} satisfy \eqref{rel}; namely,
\EQ{
\psi_1=\frac1{\sqrt2}(\psi_++i\psi_-)\ ,\qquad \psi_2=\frac1{\sqrt2}(\psi_+-i\psi_-)\ .
}
Using this fundamental system we can write down the inverse of the uniformisation map as in \eqref{ab2}. We can define coordinates in the cover that vanish at the image of each of the branch points $w=a_j$ via
\EQ{
W_j(w)=\frac{\psi_1(w)-\xi_j\psi_2(w)}{-\bar\xi_j\psi_1(w)+\psi_2(w)}\ ,
\label{ucc}
}
where
\EQ{
\xi_j=\frac{e^{2t(a_j)}+i}{e^{2t(a_j)}-i}\ .
}
\begin{figure}[ht]
\begin{center}
\begin{tikzpicture} [scale=0.7,every node/.style={scale=0.8},decoration={markings,mark=at position 0.5 with {\arrow{>}}}]]
\draw[very thick,fill=black!30,opacity=0.2] (0,0) circle (4cm);
\draw[white,fill=white] (4,0) circle (1cm);
\draw[white,fill=white] (-4,0) circle (1cm);
%\draw[white,fill=white] (4,-0.6) circle (1cm);
%
\draw[thick,red] ([shift=(0:1cm)]2.9,1) arc (95:180:1cm);
\draw[thick,red, dotted] ([shift=(0:1cm)]2.9,1) arc (95:0:1cm);
\draw[thick,red] ([shift=(0:-1cm)]-2.9,1) arc (85:0:1cm);
\draw[thick,red, dotted] ([shift=(0:1cm)]2.9,1) arc (95:0:1cm);
\draw[thick,red, dotted] ([shift=(0:-1cm)]-2.9,1) arc (85:180:1cm);
%\draw[thick,red,dotted] ([shift=(-32:1cm)]4,0.6) arc (-32:-158:1cm);
%
\draw[thick,blue] ([shift=(0:1cm)]2.9,-1) arc (-95:-180:1cm);
\draw[thick,blue, dotted] ([shift=(0:1cm)]2.9,-1) arc (-95:0:1cm);
\draw[thick,blue] ([shift=(0:-1cm)]-2.9,-1) arc (-85:-0:1cm);
\draw[thick,blue, dotted] ([shift=(0:-1cm)]-2.9,-1) arc (-85:-180:1cm);
%\draw[thick,blue,dotted] ([shift=(32:1cm)]4,-0.6) arc (32:158:1cm);
%
\filldraw[black] (3,0) circle (0.1cm);
\filldraw[black] (5,0) circle (0.1cm);
\filldraw[black] (-3,0) circle (0.1cm);
\filldraw[black] (-5,0) circle (0.1cm);
\draw[dotted,thick] (0,-4) -- (0,4);
\node at (1.5,1.5) {\bigcircled{1}};
\node at (-1.5,1.5) {\bigcircled{2}};
\node at (2.0,0) {${ W_1(a_2)}$};
\node at (6.2,0) {${ W_1(1/\bar{a}_2)}$};
\node at (-1.8,0) {${ -W_1(a_2)}$};
\node at (-6.5,0) {${ -W_1(1/\bar{a}_2)}$};
\node at (0,-3) {\large{ ${ \cal D}$}};
%\node at (0,-4.5) {\large{ $\partial{ \cal D}^-$}};
\filldraw[red] (0,0) circle (0.1cm);
\end{tikzpicture}
\caption{\footnotesize The fundamental domain ${\cal D}$ for 2 QES and replica number $n=2$ in the coordinate with $W_1(a_1)=0$. The images of the two branch points in the $w$-plane at $w_3=a_2$ and $ w_4=1/\bar a_2$ are depicted as solid black circles. The arcs of the circles cut out from the disc are identified with their respective images. The fundamental domain is a cylinder with 2 disjoint boundaries.}
\label{fig5} 
\end{center}
\end{figure}
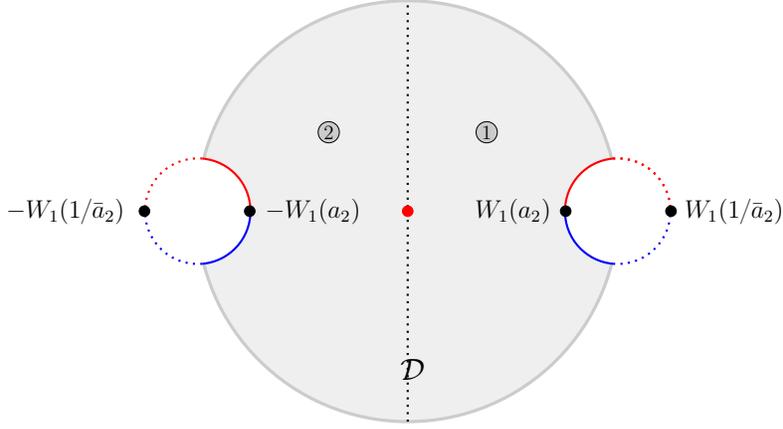
The exact expression for the accessory parameters can be computed from the poles of the Liouville stress tensor ${\EuScript T}(w)$ in \eqref{der}
\EQ{
c_j=-\frac{6\sum_kw_{jk}^2-\sum_{kl}w_{kl}^2+32\alpha^2}{32\prod'_kw_{jk}}\ ,
\label{xox2}
}
where $c_j=(c^-_1,c^+_1,c_2^+,c_2^-)$ and the product in the denominator excludes the term with $k=j$. In addition, the coefficients $\lambda_j$ in the expansions \eqref{zee} of the covering space coordinates around a branch point, are
\EQ{
\lambda_j=-\frac{4\alpha e^{2t(a_j)}(e^{2\bar t(a_j)}+i)}{e^{2t(a_j)}-e^{2\bar t(a_j)}}\frac1{\sqrt{\prod'_kw_{jk}}}\ .
}
\subsection{Map to the torus}
For the genus one replica geometry  one can, of course,  directly construct a map from the $w$-plane to the torus through the elliptic parametrisation in terms of the Weierstrass $\wp$-function,
\bea
{\wp}(z) = \frac{h'(w_4)}{4(w-w_4)}+\frac{h''(w_4)}{24}\,,\qquad   \qquad h(w)\equiv\prod_{j=1}^4(\tilde w-w_j)\,.
\eea
\begin{figure}[h]\begin{center}
\begin{tikzpicture} [scale=0.7, decoration={markings,mark=at position 0.5 with {\arrow{>>}}}]

%decoration = {snake,   % <-- added
 %                   pre length=1pt,post length=3pt,% <-- for better looking of arrow,
 %                   }]
%
%
%
\draw[black,thick] (-6.0,0.0) -- (6.0,0.0) -- (4.0,-5.0) --(-8.0,-5)--cycle;
\draw[-stealth, black] (3.5, -5)--(4.0, -5);
\draw[-stealth, black] (-6.2, -0.5)--(-6, 0);
\draw[blue, very thick, fill = blue!10] (-4.0,0.0) -- (3.50,0.0) -- (1.50,-5.0) --(-6.0,-5)--cycle;
\draw[-> ,dashed, very thick, blue!80] (0,-5.0) -- (2,0);
\draw[-> ,dashed, very thick, blue!80] (-7.6,-4.0) -- (4.4,-4);
\draw[decorate , thick, blue!80] (-1.5,0) -- (-1,0);
\draw[decorate , thick, blue!80] (-3.5,-5) -- (-3,-5);
%\draw[-stealth, dashed, thick, blue!80] (-2,-5.0) -- (0,0);
 %   \draw[very thick,rotate around={45:(1,1.5)},red] (1,1.5) ellipse (50pt and 20pt);
%    \draw[very thick,rotate around={45:(3,1.5)},red] (3,1.5) ellipse (50pt and 20pt);
%    \draw[very thick,rotate around={45:(-1,2.15)},red] (-1,2.15) ellipse (50pt and 20pt);
 %   \draw[-stealth,decorate, very thick, blue!80] (2,2.7) -- (0,0.3);
%\draw[-stealth,decorate, very thick, blue!80] (4,2.7) -- (2,0.3);
%\draw[-stealth,decorate, very thick, blue!80] (0,3.5) -- (-2,0.8);
%\draw (0,0) ellipse (10,5);
\node at (3,-3.5) {$A$};
\node at (1.6,-2.5) {$B$};
\node at (-7.1, 0.0) { $2{\bm\omega}_2$};
\node at (5, -5.0) {$2{\bm \omega}_1$};
\node at (-3.5, -0.5) { ${{ {\cal D}}}$};
\filldraw[red] (-7,-2.5) circle (0.15cm);
\filldraw[red] (-1,-2.5) circle (0.15cm);
\filldraw[red] (-2,-5) circle (0.15cm);
\node at (-2, -5.8) {${\bm \omega_1}$};
\node at (-7.9, -2.5) {${\bm\omega_2}$};
\node at (-1, -3.3) {${\bm\omega_3}$};
\end{tikzpicture}
\caption{\footnotesize {The two-sheeted replica geometry with two branch cuts, can be mapped to the torus with periods $(2{\bm\omega}_1, 2{\bm\omega}_2)$ represented as the fundamental parallelogram with opposite edges identified. The shaded region in blue is the image of the fundamental domain ${\cal D}$ for $n=N=2$ on the torus, with the $A$- and $B$-cycles of the homology basis shown, the latter lying entirely inside ${\cal D}$. The replica wormhole geometry is thus a cylinder connecting two boundaries. } }
\label{torus}
\end{center}
\end{figure}
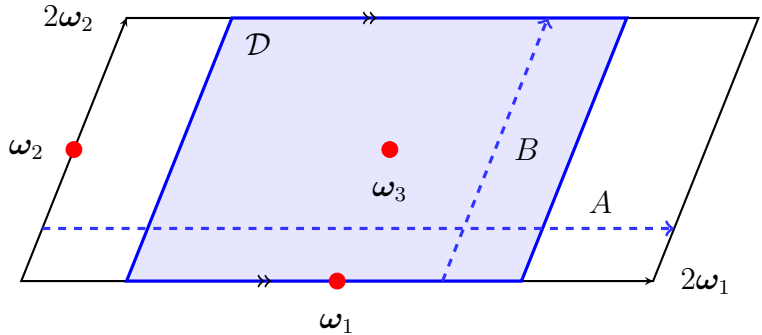
 This transformation maps the branch points $\{w_i\}_{i=1\ldots 4}$  to distinguished points on the  parallelogram with period $(2\bm\omega_1, 2\bm\omega_2)$ (see figure \ref{torus}):\footnote{The two periods of the torus are determined in terms of the branch point locations $\{w_i\}$ as $2{\bm\omega}_1=-i\pi/\alpha$ and $2{\bm\omega}_2 = \pi K(1-x)/(\alpha K(x))$, where $\alpha$ is given by \eqref{alpha}. }
 \be
 z(w_1) = \bm\omega_2\, \qquad z(w_2) =\bm\omega_1+\bm\omega_2\,\qquad z(w_3) = \bm\omega_1\qquad z(w_4)=0\ .
 \ee
 The fundamental domain ${\cal D}$ maps to a strip along the $B$-cycle of the torus, which is a finite cylinder with two boundaries. In the elliptic parametrisation $t(w)$  simply yields a point on the torus and is identified, upto a multiplicative constant, with the uniformised coordinate on the torus:
 \be
 t(w)= -\alpha\,(z- z_{-1})\,,\qquad z_{-1}\,\equiv\,z(-1)\,,
 \ee
where the constant $\alpha$ is fixed as in eq. \eqref{alpha}. The complex structure parameter of the torus is 
\be
\bm\tau\equiv\frac{\bm\omega_2}{\bm\omega_1} = i \frac{K(1-x)}{K(x)}\,,
\ee
which diverges in the small $x$ limit as,
\be
\bm\tau \big|_{x\ll 1} \approx -\frac{i}{\pi}\log x\,,
\ee
so the $A$-cycle shrinks to zero in this regime. The metric on the fundamental domain in uniformised coordinates is particularly simple:
\be
ds^2 = \frac{dt \,d\bar t}{4 \sin^2(2\,{\rm Im}\, t)}\,,\label{trumpet}
\ee
and the two disjoint boundaries of ${\cal D}$ correspond to ${\rm Im} \,t=0$ and  ${\rm Im} \,t=\pi/2$, a strip of width set by the  $A$-cycle  half-period ${\bm \omega}_1$. This is  precisely the double trumpet wormhole geometry  (see eg.~\cite{Garcia-Garcia:2020ttf}).

\subsection{Factorization limit and late times}

In the case of two QES  when the cross ratio  in \eqref{fus} $x\ll1$, we approach the factorization limit. In the Euclidean set up, this happens when $a_j$ and $1/\bar a_j$ are becoming close, i.e.~$|a_j|\to1$ and $|a_j-a_k|$ is fixed.

 In Lorentzian signature, the conformal cross-ratio as a function of the light-cone coordinates of the QES is,
 \be
 x= \frac{(a^+_1 a^-_1 +1)(a^+_2 a^-_2 +1)}{(a^+_1 a^-_2 +1)(a^-_1 a^+_2 +1)}\,.\label{fus2}
 \ee
The configuration with 2 QES naturally contributes to the generalised entropy computation of the disjoint union of two semi-infinite radiation intervals prepared  in the thermofield double state \cite{Hollowood:2021wkw}, and in terms of the bath coordinates,
\be
a_1^\pm = \mp e^{\pm2\pi/\beta (-t \pm y_1)}, \qquad a_2^\pm = \pm e^{\pm2\pi/\beta (t \pm y_2)}\,.
\ee
factorization  is naturally achieved at {\em late times} when $x \sim {\mathscr O}(e^{-4\pi t/\beta})$, and when the replica wormhole saddle is also clearly the dominant contribution to the R\'enyi entropy.

With the explicit integral expression for the map $W(w)$ in \eqref{ucc}, we find that for $w$ in the neighbourhood of, say $a_1$ and $1/\bar a_1$, as expected, the map approximates the single QES situation \eqref{lak} with $n=2$. In this limit, the accessory parameters have an expansion in powers of $x$, around the $N=1$ limit \eqref{accbasic},
\EQ{
c^\pm_j=\pm\frac{3\bar a_j}{8(1-|a_j|^2)}\big(1-\tfrac{1}{16}x^2 + {\mathscr O}(x^3)\big)\ .
\label{ubb}
}

\subsection{The fundamental domain in the factorization limit}
What is the fate of the fundamental domain ${\cal D}$ in the factorization limit $x\ll1$?

To answer this, it is useful to examine the two coordinates $W_1$ and $W_2$ when $x\ll 1$. These are  defined via $W_j(a_j)=0$, and related by 
\EQ{
W_1=\frac{(1-\xi_1\bar\xi_2)W_2+\xi_2-\bar\xi_1}{(\bar\xi_2-\xi_1)W_2+1-\bar\xi_1\xi_2}\ .
}
This is an explicit expression for the $\text{SU}(1,1)$ transformation in \eqref{rsu}. It follows that the images of the branch points $w=a_j$ in the covering coordinates $ W_{k\neq j}$ are 
\EQ{
W_1(a_2)=\frac{v}{\bar u}=\frac{\xi_2-\bar\xi_1}{1-\bar\xi_1\xi_2}\ ,\qquad W_2(a_1)=-\frac vu=-\frac{\xi_2-\bar \xi_1}{1-\xi_1\bar\xi_2}\,.
}
The parameters have nontrivial scaling in the factorization  limit $x\to 0$. For example,  with $a_i$ real, $a_1\to -1$, and $a_2$ fixed,  we find $|e^{t(a_2)- t(a_1)}| \sim x^{-1/2}$\,,
\EQ{
|W_1(a_2)|=|W_2(a_1)|=1-{\mathscr O}(x).
\label{eec}
} 
More generally for any $n$, given  a choice of coordinate $W_i$, with a particular QES $a_i$ being at the origin in the covering space, the images of the other QES $a_{j\neq i}$ are approaching the boundary of the disc $|W_i|=1$ in the small $x$ regime. Hence, in this limit, the associated circles of the Fuchsian uniformisation in figures \ref{fig2} and \ref{fig5} become very small, \footnote{More precisely, the radii are $|2uv|^{-1}\sim x$. We also note that the coefficients $\lambda_1\sim 1/x^{3/2}$  and $\lambda_2\sim {1}/{\sqrt x}$ as $x\to 0$, taking $a_1\to 1/\bar a_1$ keeping $a_2$ fixed, and vice versa if we swap the roles of $a_1$ and $a_2$. } and the fundamental domain ${\cal D}$ for both choices of coordinate, is essentially the whole of the unit disc. 
\subsection{Off-shell gravity action}
The  basic form of the gravity action is formally easy to determine in terms of the accessory parameters and the (undetermined) boundary map $w(\tau)$, as this follows from the singularities of the Liouville stress tensor \eqref{der}. For general $n$ it is,
\EQ{
&I_{\rm JT}^{(n)}[w; a_j, \bar a_j]=-NS_0 -\frac{n\phi_r}{4\beta G_N}\int_{|w|=1}d\tau\{w,\tau\}\\[5pt] &\qquad-\frac{i\pi n\phi_r}{\beta G_N}\sum_j\Big\{\varepsilon_n\partial_w\dot w_+(a_j)+c^+_j\dot w_+(a_j)-\varepsilon_n\partial_w\dot w_-(1/\bar a_j)-c^-_j\dot w_-(1/\bar a_j)\Big\}\ .
\label{gra}
}
The main obstacle to computing this precisely is (lack of)  knowledge of the boundary map $w(\tau)$ which is determined by the energy balance equation across the interface \eqref{ebe} that is generally quite complicated to solve via the welding problem.  But we can make good progress in the small $\kappa$, high temperature limit \eqref{kappa} and work at leading order so that,
\be
w(\tau)=e^{i\tau} + {\mathscr O}(\kappa)\,,\qquad \dot w (w) = iw+ {\mathscr O}(\kappa)\,.
\ee
As per our conventions in \eqref{analytic}, then $\dot w_+= i w$ and  $\dot w_-= 0$ so,
\be
I_{\rm JT}^{(n)}\big|_{\kappa\to 0}=-NS_0 +\frac{n\pi\phi_r}{\beta G_N}\left(\sum_{j=1}^N c_j^+ a_j^+ + N\varepsilon_n-\tfrac14\right)\, \label{sjt2}.
\ee
The off-shell QES dependence of the gravity action is completely determined by the accessory parameters which depend non-trivially on the $\{a_j\}$ and $\{\bar a_j\}$ as in \eqref{xox2} for $n=2$ with $N=2$.

\subsection{Dilaton dependence:  a first pass}
Although it is easy to write down the explicit dependence of the gravity action on the positions of the branch points, it is less obvious how the action can be re-expressed in terms of dilaton. In particular, how does the area  contribution to generalised R\'enyi entropies appear when there is more than one branch point ?

To gain some intuition, it is useful to express the gravity action in terms of the dilaton function evaluated at the branch point (e.g. \eqref{dilgen}), when there is only one QES i.e. $N=1$:
\be
 {\phi}_{1}(a, \bar a)\equiv \frac{2\pi \phi_r}{n\beta } \frac{1+ a \bar a}{1-a \bar a}\,.
\ee
We will treat $a$ and $\bar a$ as independent variables, which is the appropriate  viewpoint, particularly upon analytic continuation to Lorentzian light-cone variables. Then, using \eqref{xox2} and \eqref{sjt2}, we find that the gravity action for $n=2$, at leading order in the small $\kappa$ limit, with two branch points is,
\bea
&&I_{\rm JT}^{(2)}=-2S_0+\frac32\cdot\frac{1}{4G_N}\Big(\phi_{1}(a_1,\bar a_1)+\phi_{1}(a_2,\bar a_2){\Big)}\big(1- {\cal K}(x)\big)\,,\\\nonumber\\\nonumber
&&{\cal K}(x)\,\equiv\,\frac{1}{12}\left(-\frac{\pi^2}{ K(x)^2}+ 4-2x\right)\,=\, \frac{x^2}{32} + \frac{x^3}{64}+\ldots\,.
\eea
$x$ is the cross-ratio defined in \eqref{fus}, which, upon continuation to Lorentzian signature, is replaced by \eqref{fus2}. In the factorization limit when $x\to 0$, we recover the expected dependence of the action as  the sum of dilaton contributions, with the correct normalisation, evaluated at each of the branch points \eqref{pey}. 
The corrections to this expression are encoded in ${\cal K}(x)$. Coefficients in the expansion of  ${\cal K}(x)$ around $x=0$ are all positive and the function is monotonic  in the domain $0\leq x<1$. 

Given that we have a closed form expression for  the off-shell gravity action for all $x$, it would be very interesting to apply the formula to compute the generalised $n=2$ modular entropy in the time evolving evaporating black hole setup.  In section \ref{pagetimes}, we will perform a simpler analysis in the factorised limit to compute the late time value of the generalised R\'enyi entropy and its Page time as a function of the replica number $n$.

Whilst the gravity action can be determined completely as an off-shell function of QES locations,  to understand how the generalised R\'enyi  entropy depends on the area functional, we need the dilaton evaluated at the branch points in the $N=2$ situation. This should be obtained by solving the dilaton equation of motion \eqref{yfr} in the double trumpet background \eqref{trumpet}, written in terms of the boundary curves, generalising \eqref{jjw1}. We leave a detailed study of this to future work.
Below we will explain how this works for general branch point number $N$ and any $n$, and provide nontrivial confirmation of the intuitive picture above: a simple generalised modular entropy  appears naturally in a factorization limit, corrections to which are computable and encode the backreaction of branch points/cosmic branes on each other.

It is interesting to note that  for the second R\'enyi entropy, cross-ratio \eqref{fus} can also be expressed in terms of the $N=1$ dilaton as,
\be
x= \frac{\phi_{1}(a_1, \bar a_2)+\phi_{1}(a_2, \bar a_1) }{\phi_{1}(a_1, \bar a_1)+\phi_{1}(a_2, \bar a_2) }\,.
\ee
This yields a complicated, nonlinear dependance on the naive $(N=1)$ area functional, encoding interactions between the branch points. 

\section{Liouville theory and the accessory parameters}

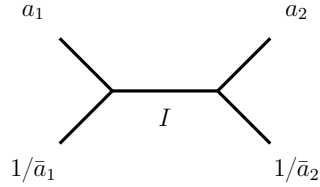
\begin{figure}[ht]
\begin{center}
\begin{tikzpicture}  [scale=0.7,every node/.style={scale=0.8}]
\draw[very thick] (0,0) -- (2,0);
\draw[very thick] (-1,-1) -- (0,0) -- (-1,1);
\node at (-1.5,-1.5) {$1/\bar a_1$};
\node at (-1.5,1.5) {$a_1$};
\draw[very thick] (3,-1) -- (2,0) -- (3,1);
\node at (3.5,-1.5) {$1/\bar a_2$};
\node at (3.5,1.5) {$a_2$};
\node at (1,-0.5) {$I$};
\end{tikzpicture}
\caption{\footnotesize The fusion channel for the Liouville correlator. The internal line with the label $I$ denotes  the descendants  of the identity operator.}
\label{fig4} 
\end{center}
\end{figure}

For general numbers of replicas and branch points, the accessory parameters are not known. However, there is a way to relate them to a particular conformal block of quantum Liouville theory in an appropriate classical limit \cite{zam2, Perlmutter:2015iya, Zam1, Harlow:2011ny, Hartman:2013mia, Litvinov:2013sxa}. The block describes a $2N$-point correlation function with insertions at the points $w_j$, collectively the set of all the pairs $(a_j,1/\bar a_j)$, with operators of dimension $h_j=\varepsilon_n c/6$. The fact that the monodromy problem has $M_{\bar j}M_j=I$ means that the block has the identity operator and its descendants (the Virasoro channel) on each of the internal lines as shown for the case $N=2$ in figure \ref{fig4}. This completely specifies the block ${\cal F}(w_j;h_j=\varepsilon_n c/6,h_a=0)$, where the $h_a$ are the dimensions of the highest weight representations exchanged on the internal lines. In the classical limit $c\to\infty$, there is a notion of a classical conformal block defined via
\EQ{
f(w_j) =\lim_{c\to\infty}\,-\frac 6c\log{\cal F}(w_j;\varepsilon_n c/6,0)\ .
}
The accessory parameters are then simply given by differentiating the classical block:
\EQ{
c_{w_j}=-\partial_{w_j}f(w_j)\ ,
\label{acc}
}
with $c_{w_j}\equiv c^+_j$ for $w_j=a_j$ and $c_{w_j}\equiv c^-_j$ for $w_j=1/\bar a_j$.

For example, for the case with 2 branch points $a_j$, $j=1,2$, the correlator is a 4-point function and in general we can write ${\cal F}(w_j)=w_{21}^{-2h}w_{43}^{-2h}\tilde{\cal F}(x)$ where $x$ is the cross ratio \eqref{fus}. This gives the classical block in the form\footnote{The function $\tilde f(x)$ is often defined to have a term $2\varepsilon_n\log x$. We have included this in the first term.}
\EQ{
f(w_j)=2\varepsilon_n\log(w_{21}w_{43})+\tilde f(x)\ .
\label{imb}
}
The block can naturally be expanded in the cross ratio \eqref{fus} and remarkably $\tilde f(x)$ can be determined to arbitrary order in the auxiliary expansion parameter $q$ defined in \eqref{defq} via a recursion relation \cite{Zam1} (see also \cite{Hartman:2013mia,Perlmutter:2015iya}). Expanding the first few term of the $q$ expansion in powers of $x$ gives 
\EQ{
\tilde f(x)=-\frac{\varepsilon_n^2x^2}3-\frac{\varepsilon_n^2x^3}3+\frac{\varepsilon_n^2}{270}(11\varepsilon_n^2-6\varepsilon_n-81)x^4+{\mathscr O}(x^5)\ .
}
For the case $n=2$, we can compare the exact expression for the accessory parameter in \eqref{xox2} with the expansion in $x$ above and find exact agreement.  At leading order in the small $x$ expansion, the accessory parameter is determined by the the term $\sim (\log w_{21}w_{43})$ in \eqref{imb},
\EQ{
c_j^\pm=\pm\frac{2\varepsilon_n\bar a_j}{1-|a_j|^2}+{\mathscr O}(x)\ .\label{lob}
}
as expected for the single QES problem. The connection with Liouville theory also provides explicit expressions for the coefficients $\lambda_j$ in \eqref{zee} \cite{Hadasz:2006rb}. 

%In order to write the expression, we need to allow $\varepsilon_n$, the coefficients of the double poles of ${\EuScript T}(w)$ in \eqref{der} to be distinct $\varepsilon_n\to\varepsilon_{1/\xi_j}$. Then define $\varepsilon_{1/\xi_j}=(1-\xi_j^2)/4$ and at the end set $\xi_j=1/n$. In this generalization, the Liouville correlator involves fields of dimension $h_j=\varepsilon_{1/\xi_j}c/6$ and so the classical conformal block now has implicit dependence $f(w_j;\varepsilon_{1/\xi_j})$. It can then be shown that \cite{Hadasz:2006rb},
%\EQ{
%\lambda_j=\frac1{2\xi_j}\exp\Big[\partial_{\xi_j}f(w_j;\varepsilon_{1/\xi_j})%\Big]_{\varepsilon_{1/\xi_j}=\varepsilon_n}\ .
%}
%In the factorization limit, the leading order behaviour is given by $f(w_j;\varepsilon_{1/\xi_j})=2\varepsilon_1\log w_{12}+2\varepsilon_2\log w_{34}$, which determines the behaviour
%\EQ{
%\lambda_1\thicksim x^{1/2n}\ ,\qquad \lambda_2\thicksim x^{-1/2n}\ ,
%\label{zol}
%}
%which matches the behaviour we found in section \ref{s2.2} for the case $n=N=2$.

\section{The dilaton}\label{s4}

In this section, we turn to the evaluation of  the dilaton at branch points when $N>1$. The dilaton at an arbitrary point in the covering space is expressed as a contour integral over the boundary of the disc as we quoted earlier \eqref{jjw1} for the case $N=1$. As explained in Appendix \ref{app:dilaton}, when $N>1$, this expression is not the complete story since the fundamental domain ${\cal D}$ has partial circles cut out of it reflecting  additional cycles with associated gluing conditions. However, in the factorisation limit $x\to 0$, when these partial circles are parametrically small, we can continue to make use of the $N=1$ expression \eqref{jjw1} as a systematic  approximation to the late time limit.

Let us  evaluate the integral \eqref{jjw1} by picking up residues as in the $N=1$ case in section \ref{s2}. Note that the derivative $\dot W=\partial_wW\dot w$ is to be thought of as a function on the cover. 
The next step to facilitate the evaluation of the integral, is to split $\dot w(w)$ as in \eqref{spl} where $\dot w_\pm(w)$ are analytic inside/outside the unit disc $|w|=1$. 

Proceeding as for $N=1$, to evaluate the integral \eqref{jjw1}, we pull the contour off to 0 for $\dot w_+$ and $\infty$ for $\dot w_-$. There are poles of order 2 coming from the denominator and poles of order $n-1$ from the images of the branch points $W(a_j)$ inside the circle and $W(1/\bar a_j)$ outside the circle. The new element in the discussion is that there will be additional poles coming from the circular regions that are cut out to define the fundamental domain ${\cal D}$ of the uniformisation, as shown in figure \ref{fig2}. 

Firstly, consider the contribution of the double poles coming from from the denominator of \eqref{jjw}. For $\tilde W=W$ and $1/\bar W$, respectively, we have a contributions
\EQ{
\phi_0^+&=-\frac{2\pi i\phi_r}\beta\Big(\frac{2\bar W\dot W_+(W)}{1-| W|^2}+\partial_W\dot W_+( W)\Big)\ ,\\[5pt]
\phi_0^-&=-\frac{2\pi i\phi_r}\beta\Big(\frac{2\bar W\dot W_-(1/\bar W)}{1-| W|^2}-\partial_{W}\dot W_-(1/\bar W)\Big)\ .
\label{nns1}
}

Now we turn to the contribution from one of the images of the branch points, e.g.~one inside the disc $\tilde  W=W(a_k)$, coming from the factor $\partial_wW$. It is simpler to work in the coordinate $W_k$ that vanishes at $w=a_k$. We denote our arbitrary point as $(W_k,\bar  W_k)$ in this coordinate.
The behaviour around the branch point is
\EQ{
\dot W_+=\frac{\lambda_k^n(W_k)^{1-n}}n\Big\{\dot w_+(a_k)+\big(\varepsilon_n^{-1}c^+_k\dot w_+(a_k)+\partial_w\dot w_+(a_k)\big)(W_k/\lambda_k)^n+\cdots\Big\}\ .
\label{zea}
}
Picking up the residue coming from the first term in the expansion \eqref{zea}, gives a contribution
\EQ{
\phi_k^+=-\frac{2\pi i\phi_r}{n\beta}(1-| W_k|^2)^2\lambda_k^nd_k\dot w_+(a_k)\ ,
\label{pi1}
}
where
\EQ{
d_k&=\oint_0\frac{dy}{2\pi i}\,\frac{y^{1-n}}{( W_k-y)^2(1-y\bar W_k)^2}\\[5pt]
&=\frac{ W_k^n(n(| W_k|^2-1)-| W_k|^2-1)+\bar W_k^n(n(| W_k|^2-1)+| W_k|^2+1)}{(| W_k|^3-1)^3}\ .
}
There is a similar contribution from rom the branch point $\tilde W=W(1/\bar a_k)$. The behaviour of the coefficient $d_k$ is important for our analysis. First of all, $d_k$ vanishes when $n=0,\pm1$ and has the following expansion around $n=1$:
\EQ{
d_k=(n-1)\frac{| W_k|^2-1-2| W_k|\log| W_k|}{ W_k(| W_k|-1)^3}+{\mathscr O}(n-1)^2\ .
\label{ex1}
}
On the other hand, for general $n$, $d_k$ has the following expansion in $1-| W_k|$:
\EQ{
d_k=\frac{n(n^2-1)}{6}+{\mathscr O}(1-| W_k|)\ .
\label{ex2}
}

In general, we have an expression that consists of a sum of the contribution from the double poles and multiple images of the QES. This is rather difficult to handle but actually what we really want is the dilaton evaluated at one of the QES. In this case, each double pole merges with one of the images, so for $\tilde W=W$ with 
$\tilde W=W(a_j)$, and, although separately divergent, together they give a finite contribution
\EQ{
\phi_0^++\phi_j^+=-\frac{2\pi i\phi_r}{n\varepsilon_n\beta}\Big(c_j^+\dot w_+(a_j)+\varepsilon_n\partial_w\dot w_+(a_j)\Big)\ ,
}
which comes from the subleading term in \eqref{zea}. There is a similar expression for the poles at $1/\bar W$ and  $W(1/\bar a_j)$
\EQ{
\phi_0^-+\phi_j^-=-\frac{2\pi i\phi_r}{n\varepsilon_n\beta}\Big(c_j^-\dot w_-(1/\bar a_j)-\varepsilon_n\partial_w\dot w_-(1/\bar a_j)\Big)\ .
}

Let us summarize what we have established do far. The dilaton evaluated at the QES $(a_j,\bar a_j)$ (we use coordinates in the base here) can be expressed in terms of the boundary map as 
\EQ{
\phi(a_j,\bar a_j)=-\frac{2\pi i\phi_r}{n\varepsilon_n\beta}\Big(c_j^+\dot w_+(a_j)+\varepsilon_n\partial_w\dot w_+(a_j)+c_j^-\dot w_-(1/\bar a_j)-\varepsilon_n\partial_w\dot w_-(1/\bar a_j)\Big)+\Delta\phi_j\ .
\label{ccu}
}
The final term here schematically encodes the contribution from all the other images of the branch points in the covering space $|W|\leq1$ in or on the boundaries of the uniformisation circles. Although this correction spoils the potential simplicity of the expression for the dilaton evaluated at the branch points, importantly, we expect it to be suppressed in the  small $x$ factorization limit.

For the case of a single branch point the additional corrections $\sim \Delta\phi_j$ are, of course, absent because there is only one image $W(a)$ in the disc. In this case, the accessory parameter is exactly \eqref{lob} and the expression \eqref{ccu} yields \eqref{diln1}.

\subsection{The gravitational action}

As noted previously, the gravitational action \eqref{rff} consists of two pieces. The second part contains the meromorphic function ${\EuScript T}(w)$ and can be computed using Cauchy's residue theorem after expressing $\partial_\tau w$ as a function of $w$ with the decomposition $\dot w(w)=\dot w_+(w)+\dot w_-(w)$, which yields \eqref{gra}.
%\EQ{
%I_\text{JT}[w;a_j,\bar a_j]&=-\frac{n\phi_r}{4\beta G_N}\int d\tau\,\{w,\tau\}-\frac{i\pi n\phi_r}{\beta G_N}\sum_j\Big\{\varepsilon_n\partial_w\dot w_+(a_j)+c^+_j\dot w_+(a_j)\\[5pt]&\qquad\qquad-\varepsilon_n\partial_w\dot w_-(1/\bar a_j)-c^-_j\dot w_-(1/\bar a_j)\Big\}\ .
%\label{gra}
%}
If we compare this to the expression for the dilaton evaluated at the branch points \eqref{ccu}, we immediately have
\EQ{
I_\text{JT}[w;a_j,\bar a_j]=-NS_0+\frac{n^2-1}{8G_N}\Big(\sum_j\phi(a_j,\bar a_j)+\Phi_n\Big)-\frac{n\phi_r}{4\beta G_N}\int d\tau\,\{w,\tau\}\,,
\label{ukk}
}
where the $\Phi_n$ term collects all the additional contributions $-\sum_j\Delta\phi_j$.
This is our main result since it relates the gravitational action to the dilaton evaluated at the QES. The dependence on the $a_j$ is not only in the explicit positions at the which the dilaton is evaluated but there will also be implicit dependence when the boundary map is put on shell.

\section{Eternal black hole and Page times for R\'enyi entropies }
\label{pagetimes}
\begin{figure}[ht]
\begin{center}
\begin{tikzpicture} [scale=0.6,every node/.style={scale=0.8}]
\draw[yellow!20,fill=yellow!20] (-4,-4) rectangle (4,4);
\draw[thick,fill=pink!20] (0,0) circle (2.5cm);
\filldraw[black] (1,0.7) circle (2pt);
\filldraw[black] (2.1,1.5) circle (2pt);
\filldraw[black] (-1,-0.7) circle (2pt);
\filldraw[black] (-2.1,-1.5) circle (2pt);
\draw[thick] (-1,-0.7) to[out=10,in=-170] (1,0.7);
\draw[thick] (2.1,1.5) to[out=0,in=180] (4,2);
\draw[thick] (-2.1,-1.5) to[out=180,in=0] (-4,-2);
\node at (1,0.3) {$a_1$};
\node at (2.8,1.2) {$b_1$};
\node at (-1,-0.3) {$a_2$};
\node at (-2.8,-1.2) {$b_2$};

\begin{scope}[xshift=12cm]
\filldraw[pink!20] (-3,-4) rectangle (3,4);
\filldraw[yellow!20] (6,0) -- (3,3) -- (3,-3) -- (6,0);
\filldraw[yellow!20] (-6,0) -- (-3,3) -- (-3,-3) -- (-6,0);
\draw[-] (-3,-3) -- (-6,0) -- (-3,3);
\draw[-] (3,-3) -- (6,0) -- (3,3);
\draw[-] (-3,-3) -- (3,3);
\draw[-] (-3,3) -- (3,-3);
\draw[-] (3,-4) -- (3,4);
\draw[-] (-3,-4) -- (-3,4);
\filldraw[black] (3,1) circle (2pt);
\filldraw[black] (1.6,0.8) circle (2pt);
\filldraw[black] (-3,1) circle (2pt);
\filldraw[black] (-1.6,0.8) circle (2pt);
\draw[thick] (-1.6,0.8) to[out=-10,in=-170] (1.6,0.8);
\draw[thick] (3,1) to[out=0,in=170] (6,0);
\draw[thick] (-3,1) to[out=180,in=10] (-6,0);
\node at (0,1.1) {$I$};
\node at (4.2,0.3) {$R$};
\node at (-4.2,0.3) {$R$};
\node at (3.4,1.5) {$b_1^\pm$};
\node at (-3.4,1.5) {$b_2^\pm$};
\node at (1.6,0.3) {$a_1^\pm$};
\node at (-1.6,0.3) {$a_2^\pm$};
\end{scope}
\end{tikzpicture}
\caption{\footnotesize The set up in (left) Euclidean (right) Lorenztian signature with two QES in the context of the eternal black hole and a radiation region $R$ that covers the left and right bands, as shown. There is a point in the right Minkowski bath and one QES as shown ultimately outside the horizon of the right black hole in the Lorentzian picture and the mirror image on the left.}
\label{fig6} 
\end{center}
\end{figure}
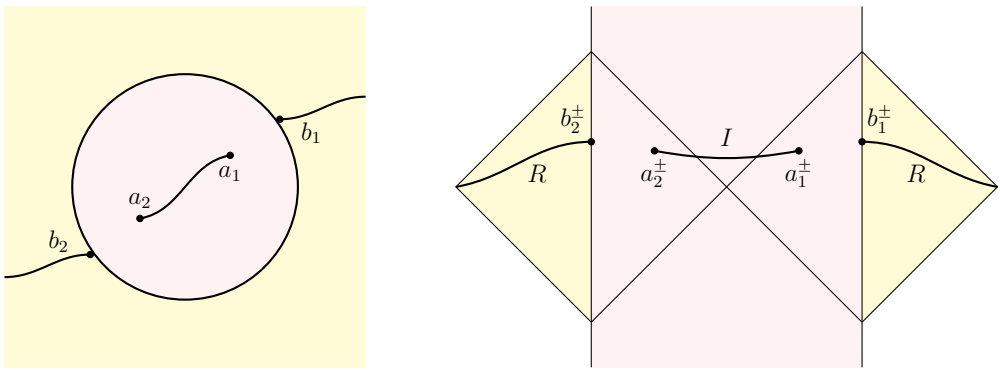

As a concrete application we consider the case of the eternal black hole with two QES as shown in figure \ref{fig6}. Once continued to Lorentzian signature, this configuration describes the island saddle of a radiation region that spans the left and right bath regions. We choose the end points to lie just outside the interface between the AdS region and the baths. We also take the setup to be symmetric in Lorentzian signature for simplicity, so with coordinates
\EQ{
b_1=-b^-_1=-\bar b_2=-b^+_2=e^{-t_0} \ ,\qquad \bar b_1=b_1^+=-b_2=b_2^-=e^{t_0}\ .
}
Again the Lorentzian time here is re-scaled in units of $\beta/2\pi$. The QES of the island saddle will also inherit this symmetry, so $a_1=-a^-_1=-\bar a_2=-a^+_2$ and $\bar a_1=-a^+_1=-a_2=a_2^-$.

The next task is to solve the equation of motion \eqref{ebe} to find the boundary map with the QES off shell to linear order in $\kappa$. This would be a formidable task, dealing with the multiple poles in the Liouville stress tensor ${\EuScript T}(w)$ without some kind of further approximation. We already know in the context of the $n\to1$ limit and the von Neumann entropy that the island saddle can only become dominant over the ``no-island", or Hawking, saddle in the late time $t_0 \gg 1$ limit. 

So we  work in the limit of small $e^{-t_0}\ll1$, alongside the high temperature limit
 $\kappa\ll1$, although it is important that there is a hierarchy of scales $e^{-t_0}\ll\kappa\ll1$. This separation of scales follows from the fact that $t_0$ is  of the order of the Page time or greater, when the island saddle can dominate, and $|\log\kappa|$  sets the  scrambling time (measured in units of $\beta/2\pi$).

In the late time limit, it can be seen that locations of the poles at $a_1$ and $1/\bar a_1$ will end up scaling like $b_1$ and $1/\bar b_1$, i.e.~as $e^{-t_0}$, whereas $a_2$, $1/\bar a_2$ will end up scaling like $b_2$ and $1/\bar b_2$, i.e.~as $e^{t_0}$. So the boundary map in the vicinity of $a_1$ and $1/\bar a_1$ is decoupled from the poles at $a_2$ and $1/\bar a_2$, and vice-versa. The corrections would come in at order $e^{-2t_0}$. So at lower order, from the point-of-view of the boundary map, we can essentially use the 1 QES result in the vicinity of each pair $(a_j,1/\bar a_j)$. There are also corrections of order $e^{-3t_0}$ from the expansions of the accessory parameters $c_j^\pm$ to the leading order expressions \eqref{lob}. 
There are also corrections of the same order in relating the gravitational action to the dilaton, the additional inter-QES terms $\Delta\phi_j$ in \eqref{ukk} are also of the same order $e^{-2t_0}$. Finally the CFT entropy will take the form of the sum of two decoupled expressions for the left and right black holes up to terms of order $e^{-2t_0}$.

We conclude that to order $e^{-2t_0}$, the left and right hand sides can be treated as factorized as far as the gravitational action, boundary map and CFT entropy is concerned. This is also true of the functions $F_j(w)$ and $G_j(w)$ that solve the conformal welding problem which now have an approximate form on the right, $j=1$, and the left $j=2$. These are of the form \eqref{weld} with $b\to b_j$ and $\bar b\to\bar b_j$. 

The modular entropy of the no-island saddle is equal to
\EQ{
\tilde S_n^\text{gen}(\emptyset)&=\frac{c}{6n}\log\frac{|F_1(b_1)-F_2(b_2)|^2}{|F'_1(b_1)F'_2(b_2)b_1b_2|}\\[5pt]
&=\frac{c}{3n}\Big(t_0-\frac{3\kappa(n^2-1)}{4(4n^2-1)}+\cdots\Big)\ ,
}
where the corrections involve powers of $\kappa$ and terms of order $e^{-2t_0}$. For the island saddle, we need only take twice the entropy of the island saddle in the one QES scenario in \eqref{sgen} with $|b|=1$: 
\EQ{
\tilde S_n^\text{gen}(I)=\frac{4}{n+1}S_0+\frac{c}{3n}\Big(\frac1\kappa-2\kappa-\frac{9\kappa(n^2-1)}{4(4n^2-1)}+\cdots\Big)\ .
}
Hence, the Page time, the time when the two saddles have equal entropy, occurs when
\EQ{
t_0=\frac{6\beta nS_0}{\pi c(n+1)}+\frac{\beta}{2\pi}\Big(\frac1\kappa-\frac{\kappa(19n^2-7)}{2(4n^2-1)}+\cdots\Big)\ ,
}
where we have re-instated the units $\beta/2\pi$.

\section{Discussion}\label{s5}

We have studied the computation of generalised R\'enyi entropies  in a CFT radiation bath coupled to a black hole within a dynamical JT gravity framework, using the replica trick for finite replica number $n$. Unlike the $n\to 1$ limit wherein the back-reaction of the cosmic branes or twist fields on the background and on each other can be neglected, we are now required to consider these effects carefully. 

Employing the theory of Fuchsian uniformisation we were able to obtain the general form of the action functional for the off-shell locations of the QES, in terms of accessory parameters of the associated Liouville stress tensor, and the dynamical boundary map. 
The boundary map which determines the gluing between the gravitating region and the bath  is the most technically challenging piece of the puzzle, but one which can largely be circumvented by working in the high temperature, or small scrambling time, limit (see e.g.~\cite{Almheiri:2019qdq}). Particularly noteworthy is the second R\'enyi entropy for two quantum extremal surfaces, when the underlying Riemann surface is a torus and for which the accessory parameter and unformisation map can be written in closed form.

Once the gravitational action functional for the QES locations is known, the computation of generalised R\'enyi entropies can proceed automatically by extremising the sum of the gravity and CFT partition functions on the replica geometry with new branch points at  the QES. 

However, one of our  conceptual goals  was to understand how this functional can be interpreted as a generalised R\'enyi entropy. The answer to this question depends on how the dilaton in JT gravity (which plays the role of the area) evaluated at the QES locations is related to the action functional. With one QES, and generic $n$, the answer is clean: the dependence of the gravity action on QES locations is, up to an $n$-dependent normalisation  constant, precisely proportional to the dilaton. The $n$-dependent proportionality constant is important, and leads us to unambiguously identify the correct generalised entropy for arbitrary replica number   as the generalised  modular entropy exactly as might be expected from independent holographic considerations \cite{Dong:2016fnf}. For multiple QES, the relation between the action functional and the dilaton at the branch points is more complicated. The complications however are subleading in the late time limit, precisely when island saddles dominate, wherein individual QESs decouple from each other and we recover a relatively simpler extremisation problem involving the generalised modular entropy functional for multiple QES and multiple intervals. The correction terms collectively denoted as  $\Phi_n$ in the gravitational action \eqref{ukk} are parametrically small and governed by the size of the Fuchsian uniformisation circles.

There are some immediate questions that we wish to address in future work:
\begin{itemize}
\item{We have focussed attention on the Euclidean problem in this paper. An immediate pressing question is how the approach developed here can be applied to  nonequilibrium entanglement evolution within a Lorentzian setting. Understanding the construction of  replica wormholes and the computation of time dependent R\'enyi entropy (arbitrary $n$) for the Hawking radiation from an evaporating, single-sided black hole poses interesting technical and conceptual challenges even for the single QES problem.}
\item{A key step in deducing the dilaton dependence of the gravity action on the replica geometry, is the evaluation of the dilaton at  the branch points or QES. For the single QES problem this is achieved through a representation of the dilation as an integral over the boundary, where the integrand  depends on the boundary map. A similar representation is possible on the covering space in the multiple QES situation, but since the fundamental domain ${\cal D}$ is no longer the unit disc, there are additional contributions to this integral from within the circles associated to Fuchsian uniformisation. Determining the additional contributions is not straightforward, but the two QES problem with $n=2$, offers a tractable situation where these corrections can be completely determined. In the latter case, the fundamental domain is a cylinder with two boundaries, with the double trumpet metric \eqref{trumpet} and the dilaton equation of motion should be expressible as an integral over the disjoint boundaries. It would be extremely interesting to understand the corrections to the area term in gravity action in this case, and potentially understand their relation to similar terms in the holographic context \cite{Dong:2016fnf}.
}
\end{itemize}

\acknowledgments{TJH and SPK would like to acknowledge support from STFC Consolidated Grant Award ST/X000648/1. LCP is supported by a joint STFC studentship  ST/V507143/1 and  EPSRC DTP, Faculty of Science \& Engg, Swansea U. SPK would like to thank the Fields, Gravity \& Strings  group,  Laboratoire de Physique de L' \'Ecole Normale Sup\'erieure for hospitality whilst this work was being completed.}
\vspace{1in}

{\footnotesize Open Access Statement - For the purpose of open access, the authors have applied a Creative Commons Attribution (CC BY) licence to any Author Accepted Manuscript version arising. 

Data access statement: no new data were generated for this work.}

\appendix

\section{Dilaton from the boundary curve}\label{app:dilaton}

We want to derive an expression for the dilaton in terms of the boundary curve $w(\tau)$ irrespective of the CFT stress tensor subject only to some suitable analyticity properties. The strategy is to solve the wave equation \eqref{yfr} in terms of the stress tensor with  a Green's function method in the covering space, and then to use a dispersion relation to relate the stress tensor to its value on the boundary $\partial\widetilde{\cal M}_n$. Then we will use the energy conservation equation \eqref{ebe} to eliminate the stress tensor to arrive at an integral expression for the dilaton purely in terms of the boundary curve. The steps are similar to those  in appendix C in \cite{Goto:2020wnk}, but we point out additional complications arising when there are multiple branch points, and we are working away from the $n\to 1$ limit.

\subsection{Single QES problem: $N=1$}
The single QES problem ($N$=1) offers the cleanest situation. In this case the (inverse) uniformisation map to the covering space is simplest \eqref{lak} and the fundamental domain ${\cal D}$ is the unit disc in the $W$-plane.
In order to simplify the analysis we find it easier to map the unit disc in the covering space $|W|\leq1$ to the $\text{Re}\,Z\leq0$ half of the complex $Z$ plane using the Cayley transform\,,
\EQ{
W=\frac{Z_0-Z}{Z_0+Z}\ .
}
Here, $Z_0<0$ is a real arbitrary constant. The expression for the dilaton at the point $Z=Z_0$ is given by the Green's function method as
\EQ{
\phi(Z_0,Z_0)=8\pi G_N\int^{Z_0}dX\,\frac{Z_0^2-X^2}{Z_0}\,\text{Re}\,T(X)\ ,\label{dil1}
}
where $T\equiv T_{ZZ}$. The ability to add a multiple of the homogeneous solution is hiding in the choice of the lower limit of the integral. Note that we have chosen to evaluate the dilaton at what appears to be a distinguished  point with real $Z$ but we will be able to transform this to a generic point using a suitable M\"obius transformation later.

We can then relate the real to the imaginary part of the stress tensor using the integral relation
\EQ{
\text{Re}\,T(X)= \frac1\pi\int_{-\infty}^\infty dY\,\frac{Y}{X^2+Y^2}\,\text{Im}\,T(iY)\ .
\label{disp}
}
The integral here is along the image of the boundary $\partial {\cal D}$ of the disc $|W|=1$ in the $Z$ plane, i.e. the imaginary axis. 

Next, we use the energy conservation equation \eqref{ebe} to express the energy flux $\sim {\rm Im}\, T_{ZZ}$ at the AdS$_2$ boundary in terms of the boundary Schwarzian,
\be
-2 {\rm Im} \,T_{ZZ}(iY(\tau))=\left.\frac{\phi_r}{8\pi G_N} \frac{\partial_\tau\{Z, \tau\}}{\dot Z(\tau)^2}\right|_{Z=iY(\tau)}\label{ImT}
\ee		
We assume that the boundary map $z(\tau)$ (in Euclidean signature) is monotonic and therefore invertible. Hence, we can define for $z\in\partial{\cal M}_n$
\EQ{
\dot z(z)=\frac{dz}{d\tau}\Big|_{\tau=\tau(z)}\ .
}
On the boundary  we can decompose $\dot z(z)=\dot z_+(z)+\dot z_-(z)$ and analytically continue with $\dot z_+(z)$ and $\dot z_-(z)$ analytic in $\text{Re}\,z\gtrless 0$, respectively.  
This in turn allows to  define the functions in the cover $\dot Z(Z)=\dot z\partial_zZ$ using the uniformisation map $z=\pi(Z)$. We then have the following identity,
\EQ{
\frac{\partial_\tau\{Z,\tau\}}{\dot Z^2}=\partial_Z^3\dot Z\ .
}
Substituting in for the energy flux into the dispersion relation \eqref{disp}, and \eqref{dil1}, one can perform the $X$ integral (and just take the upper limit of the integral fixing the homogeneous solution) and then integrate by parts three times in $Z$ to give
\EQ{
\phi(Z_0,Z_0)=-\frac{2\phi_rZ_0^2}\pi\int_{\partial{\cal D}}dZ\,\frac{\dot Z}{(Z^2-Z_0^2)^2}\ .
}
Finally, a M\"obius transform of $(Z_0,Z_0)$ to an arbitrary point $({\cal Z},\bar{\cal Z})$ yields:
\EQ{
\phi({\cal Z},\bar{\cal Z})=-\frac{\phi_r({\cal Z}+\bar{\cal Z})^2}{2\pi} \int_{\partial{\cal D}}dZ\,\frac{\dot Z}{({\cal Z}-Z)^2(\bar{\cal Z}+Z)^2}\,,
}
on the half plane.  Transforming back to the $W$ coordinate on the disc,
\EQ{
\phi(W,\bar{W})=-\frac{\phi_r(1-|W|^2)^2}{2\pi}\int_{\partial{\cal D}}d\tilde W\,\frac{\dot W}{({W}-\tilde W)^2(1-\bar{W}\tilde W)^2}\ .
\label{jjw}
}
\subsection{Multiple QES: $N>1$}
The uniformisation problem with multiple branch points is more complex and, as we have seen, the fundamental domain ${\cal D}$ is now a disc with  circles cut out as shown in figures \ref{fig2} and  \ref{fig5}. After identification of the arcs of the cut out circles with their images,  the boundary of the fundamental domain consists of  disjoint cycles. We can, of course, pose the dilaton Green's function problem  in the covering space with boundary $|W|=1$, with the  energy balance equation \eqref{ImT}, 
but the contributions from the regions cut out need to be accounted for. This is a complicated and subtle point, particularly the question whether additional singularities appear in these regions that will contribute.  We will postpone a precise investigation of these contributions to future work. However, we note that progress can nevertheless be made, by exploiting the limit $x\to 0$. We have shown that the partial circles associated to the Fuchsian uniformisation become parametrically small with radii scaling as $x$ in this limit, and therefore the corrections to the formula \eqref{jjw} are correspondingly negligible at late times when the saddles with QES dominate.

\end{document}